\documentclass[aps,pra,twocolumn,floatfix,showpacs,preprintnumbers,amsmath,amssymb,10pt]{revtex4-1}
\usepackage[dvips]{graphics}

\DeclareMathAlphabet{\mathsfit}{\encodingdefault}{\sfdefault}{m}{sl}
\SetMathAlphabet{\mathsfit}{bold}{\encodingdefault}{\sfdefault}{bx}{sl}

\usepackage{mathptmx}
\usepackage{psfrag}
\usepackage{graphicx}
\usepackage{bm}
\usepackage{amsmath}
\usepackage{trfsigns}
\usepackage{amssymb}
\usepackage{subfigure}
\usepackage[usenames,dvipsnames]{color}
\usepackage{dashrule}
\usepackage{textgreek}
\usepackage[english]{babel}
\usepackage{contour}
\usepackage{upgreek,bm}
\usepackage{color}
\definecolor{dred}{rgb}{.6,.0,0.}
\definecolor{dblue}{rgb}{.0,.0,0.6}

\renewcommand{\vec}[1]{\mathbf{#1}}

\newcommand{\tens}[1]{\mbox{\textsf{\textbf{#1}}}}

\newcommand{\Greektens}[1]{\contour[3]{black}{#1}}

\newcommand{\sprod}{\!\cdot\!}
\newcommand{\vprod}{\!\times\!}

\newcommand{\dif}{\mathrm{d}}
\newcommand{\mi}{\textrm{i}} 
\newcommand{\me}{\mathrm{e}}

\begin{document}

\title{Casimir-Lifshitz Force for Nonreciprocal Media\\ and Applications to Photonic Topological Insulators}

\author{Sebastian Fuchs$^{1,2}$}
\thanks{These authors contributed equally to this work.}
\author{Frieder Lindel$^{1,2}$}
\thanks{These authors contributed equally to this work.}
\author{Roman V. Krems$^2$}
\author{George W. Hanson$^3$}
\author{Mauro Antezza$^{4,5}$}
\author{Stefan Yoshi Buhmann$^{1,6}$}

\affiliation{$^1$ Physikalisches Institut, Albert-Ludwigs-Universit\"at Freiburg, Hermann-Herder-Stra{\ss}e 3, 79104 Freiburg, Germany\\
$^2$ Department of Chemistry, University of British Columbia, Vancouver, British Columbia V6T 1Z1, Canada\\
$^3$ Department of Electrical Engineering, University of Wisconsin-Milwaukee, 3200 N. Cramer St., Milwaukee, Wisconsin 53211, USA\\
$^4$ Laboratoire Charles Coulomb, UMR 5221 Universit\'{e} de Montpellier and CNRS, F-34095 Montpellier, France\\
$^5$ Institut Universitaire de France, 1 rue Descartes, F-75231 Paris Cedex 05, France\\ 
$^6$ Freiburg Institute for Advanced Studies, Albert-Ludwigs-Universit\"at Freiburg, Albertstra{\ss}e 19, 79104 Freiburg, Germany}

\date{\today}

\begin{abstract} 
Based on the theory of macroscopic quantum electrodynamics, we generalize the expression of the Casimir force for nonreciprocal media. The essential ingredient of this result is the Green's tensor between two nonreciprocal semi-infinite slabs including a reflexion matrix with four coefficients that mixes optical polarizations. This Green's tensor does not obey Lorentz's reciprocity and thus violates time-reversal symmetry. The general result for the Casimir force is analyzed in the retarded and nonretarded limits, concentrating on the influences arising from reflections with or without change of polarization. In a second step we apply our general result to a photonic topological insulator whose nonreciprocity stems from an anisotropic permittivity tensor, namely InSb. We show that there is a regime for the distance between the slabs where the magnitude of the Casimir force is tunable by an external magnetic field. Furthermore the strength of this tuning depends on the orientation of the magnetic field with respect to the slab surfaces.
\end{abstract}

\maketitle

\section{Introduction}
The Casimir force in its original meaning is an attractive force between two parallel, uncharged and conducting plates in vacuum. In quantum field theory this effect can be traced back to vacuum fluctuations of the electromagnetic field. Thus Casimir \cite{Casimir:1948} originally computed the force between two perfectly conducting plates based on vacuum field fluctuations. This force scales as $1/z^4$ where $z$ is the distance between the plates.
Here, we explore how the Casimir force is modified for nonreciprocal medium and if, in the case of topological insulators, it can be tuned by a magnetic field. This work is stimulated by recent 
progress of the experiments aimed at the study of the quantum Hall effect, topological insulators and nonreciprocal materials in general \cite{Hasan:2010}.  
The unusual optical properties of nonreciprocal materials have, to the best of our knowledge, not been taken into account in previous treatments of dispersion interactions. Yet, they may change the sign and/or the scaling of the Casimir force. The theory derived here provides a general framework for the analysis of the Casimir force in a variety of experimentally relevant setups.

We compute the Casimir force as the ground-state expectation value of the Lorentz force between two bodies characterized by the charge density and the current density \cite{Buhmann_Book_1}. The electric field and the charge density are mutually correlated since a fluctuating charge density induces fluctuating electric fields and vice versa. This process intertwining the fluctuating charge densitiy and the electric field is responsible for the occurrence of a non-vanishing net force between two bodies, namely the Casimir force. The same applies to the current density and the magnetic field. The Lifshitz approach \cite{Lifshitz:1956} considers two dielectric half spaces which show randomly fluctuating polarizations. It is noteworthy that the ground-state expectation values of the electric field, the magnetic field, the charge density and the current density vanish. In the absence of correlations, the expectation value of the Lorentz force would vanish and there would be no net force.

We apply the theory of macroscopic quantum electrodynamics (QED), which incorporates the influcence of material properties by a permittivity and a permeability \cite{Scheel:2008, Buhmann_Book_1, Dung:1998} to compute the Casimir force. The Green's tensor represents the propagator between the fluctuating noise currents and the quantized electric and magentic fields. Using this theory, the Casimir force for magnetodielectric bodies has been computed in Ref.~\cite{Raabe:2005}. The theoretical extension of macroscopic (QED) for arbitrary nonlocal and nonreciprocal linear media was carried out in Refs.~\cite{Raabe:2007, Butcher:2012}. In the case of nonreciprocal media the electric and magnetic fields are coupled in a way that violates time-reversal symmetry \cite{Butcher:2012}. In QED this means that the electric field fluctuations and their source, namely noise currents, are not interchangeable. Thus the Green's tensor violates Lorentz's reciprocity principle \cite{Lorentz:1896, Onsager:1931}.

We consider two semi-infinitely extended plates with a separation of length $L$ and the respective Green's tensor contains four contributions from an even/odd number of reflections for outgoing waves to the right/left direction \cite{Li:1994, Tomas:1995}. This result for the Green's tensor is extended to nonreciprocal materials. After deriving a general expression for the Casimir force in nonreciprocal media, we apply the result to a photonic topological insulator. Whereas the axion topological insulator, cf. Ref.~\cite{Hasan:2010, Grushin:2011, Nie:2013, Rodriguez-Lopez:2014}, couples electric and magnetic fields by a quantized axion coupling which is usually much smaller than the electric and magnetic properties of the material, the photonic topological insulator \cite{Hanson:2016} shows an anisotropic permittivity, which is responsible for the nonreciprocity of the material. To calculate the Casimir force we compute the reflection coefficients for the material with a general approach for biaxial, anisotropic magnetodielectrics \cite{Rosa:2008_2}. We also analyze the dependence of a static magnetic field on the Casimir force. In this context, the influence of the surface phonon and surface plasmon polaritons on the heat transfer has been studied for Aluminum in Ref.~\cite{Joulain:2005} and for InSb in Ref.~\cite{Moncada:2015}.

This paper is structured as follows: The basic principles and expressions of macroscopic QED for nonreciprocal media are outlined in Sec.~\ref{sec:Macroscopic QED Nonreciprocal Media}. Due to the violation of Lorentz's reciprocity principle, new definitions for the real and imaginary parts of the Green's tensor are introduced and a generalized Helmholtz equation containing a conductivity tensor is presented. Sec.~\ref{sec:Casimir Force for Nonreciprocal Media} generalizes the concept of Casimir force based on the Lorentz force for nonreciprocal media, where Lorentz's reciprocity does not hold anymore. Sec.~\ref{sec:Greens Tensor Two Surfaces} is dedicated to the derivation of the Green's tensor for two semi-infinite and nonreciprocal half-spaces. The final result is given in terms of reflection matrices comprising four types of reflection coefficients with equal and alternating polarization. Afterwards the Green's tensor is used to compute the Casimir force for this geometry. Sec.~\ref{sec:Theoretical Model of the Photonic Topological Insulator} explains the anisotropic structure of the permittivity of the photonic topological insulator and outlines the material properties of the permittivity for InSb. Moreover it provides the reflection coefficients for the photonic topological insulator. Finally, Sec.~\ref{sec:Results} shows analytical results for the Casimir force in the nonretarded limit and analyzes the retarded limit for a medium with anisotropic permittivity. In the second part the dependence of the Casimir force on the magnetic field is studied. The impact of the diagonal and off-diagonal elements of the reflection matrix on the force as well as the change of sign of the field is discussed. It is pointed out how the external magnetic field changes the main contributions of the Casimir force from surface phonon and surface plasmon polaritons to hyperbolic modes.

\section{Macroscopic QED for Nonreciprocal Media}
\label{sec:Macroscopic QED Nonreciprocal Media}
The theory of macroscopic QED, cf. Ref.~\cite{Buhmann_Book_1, Buhmann_Book_2}, incorporates material properties in terms of the macroscopic permittivity and permeability. This theory is consistent with the classical theory of macroscopic electrodynamics and satisfies Maxwell's equations, the fluctuation--dissipation theorem and free space quantum electrodynamics. In contrast to the case of a reciprocal material, time-reversal symmetry is not preserved in nonreciprocal media. To account for this, the mathematical framework has to be adjusted \cite{Butcher:2012}.\\
Lorentz's reciprocity principle for tensors, e.g. the Green's tensor $\tens{G}$, does not allow for the violation of time-reversal symmetry, which can be expressed as
\begin{equation}
\tens{G}^{\textrm{T}} \left( \vec{r}', \vec{r}, \omega \right) \neq \tens{G} \left( \vec{r}, \vec{r}', \omega \right).
\label{eq:Lorentz Reciprocity}
\end{equation}
Physically this means that in a nonreciprocal material a source at $\vec{r}'$ does not create the same field at $\vec{r}$ that a source at $\vec{r}$ would create at $\vec{r}'$. A consequence of the breaking of this essential principle is the new definition of real and imaginary parts of a tensor
\begin{align}
\begin{array}{lll}
&\Re \left[ \tens{G} \left( \vec{r}, \vec{r}', \omega \right) \right] &= \frac{1}{2} \left[ \tens{G} \left( \vec{r}, \vec{r}', \omega \right) + \tens{G}^{*\textrm{T}} \left( \vec{r}', \vec{r}, \omega \right) \right]\\[2mm]
&\Im \left[ \tens{G} \left( \vec{r}, \vec{r}', \omega \right) \right] &= \frac{1}{2 \mi} \left[ \tens{G} \left( \vec{r}, \vec{r}', \omega \right) - \tens{G}^{*\textrm{T}} \left( \vec{r}', \vec{r}, \omega \right) \right].
\label{eq:Definition Real Part Imaginary Part}
\end{array}
\end{align}
In the following, the expressions for Ohm's law, the Helmholtz equations and the electric and magnetic field terms have to be redefined to account for the peculiarities of nonreciprocal materials.\\
Ohm's law describes the linear response of matter in an external electromagnetic field and reads in frequency space
\begin{equation}
\hat{\vec{j}}_{\textrm{in}} \left( \vec{r}, \omega \right) = \int \dif^3 r' \tens{Q} \left( \vec{r}, \vec{r}', \omega \right) \sprod \hat{\vec{E}} \left( \vec{r}', \omega \right) + \hat{\vec{j}}_{\textrm{N}} \left( \vec{r}, \omega \right).
\label{eq:Ohm's Law}
\end{equation}
This term is a convolution of the conductivity tensor $\tens{Q} \left( \vec{r}, \vec{r}', \omega \right)$ and the quantized electric field $\hat{\vec{E}} \left( \vec{r}', \omega \right)$. $\hat{\vec{j}}_{\textrm{in}}$ represents the internal current density and $\hat{\vec{j}}_{\textrm{N}} \left( \vec{r}, \omega \right)$ is the noise current density. Broken reciprocity now states that $\tens{Q}^{\textrm{T}} \left( \vec{r}', \vec{r}, \omega \right) \neq \tens{Q} \left( \vec{r}, \vec{r}', \omega \right)$. By making use of Ohm's law \eqref{eq:Ohm's Law}, the continuity relation in the frequency domain combining the noise charge density $\hat{\rho}_{\textrm{in}}$ and the noise current density $\hat{\vec{j}}_{\textrm{in}}$, $\mi \omega \hat{\rho}_{\textrm{in}} \left( \vec{r}, \omega \right) = \overrightarrow{\nabla} \sprod \hat{\vec{j}}_{\textrm{in}} \left( \vec{r}, \omega \right)$, and Maxwell's equations in the frequency domain
\begin{align}
\begin{array}{lll}
&\overrightarrow{\nabla} \sprod \hat{\vec{E}} \left( \vec{r}, \omega \right) &= \frac{\hat{\rho}_{\textrm{in}} \left( \vec{r}, \omega \right)}{\epsilon_0}\\
&\overrightarrow{\nabla} \sprod \hat{\vec{B}} \left( \vec{r}, \omega \right) &= 0\\
&\overrightarrow{\nabla} \vprod \hat{\vec{E}} \left( \vec{r}, \omega \right) - \mi \omega \hat{\vec{B}} \left( \vec{r}, \omega \right) &= 0\\
&\overrightarrow{\nabla} \vprod \hat{\vec{B}} \left( \vec{r}, \omega \right) + \frac{\mi \omega}{c^2} \hat{\vec{E}} \left( \vec{r}, \omega \right) &= \mu_0 \hat{\vec{j}}_{\textrm{in}} \left( \vec{r}, \omega \right)
\label{eq:Maxwell Equations}
\end{array}
\end{align}
we find the inhomogeneous Helmholtz equation for the electric field
\begin{multline}
\left[ \overrightarrow{\nabla} \vprod \overrightarrow{\nabla} \vprod - \frac{\omega^2}{c^2} \right] \hat{\vec{E}} \left( \vec{r}, \omega \right)\\
- \mi \mu_0 \omega \int \dif^3 r' \tens{Q} \left( \vec{r}, \vec{r}', \omega \right) \sprod \hat{\vec{E}} \left( \vec{r}', \omega \right) = \mi \mu_0 \omega \hat{\vec{j}}_{\textrm{N}} \left( \vec{r}, \omega \right).
\label{eq:Helmholtz Equation}
\end{multline}
The formal solution to this inhomogeneous differential equation
\begin{equation}
\hat{\vec{E}} \left( \vec{r}, \omega \right) = \mi \mu_0 \omega \int{\dif^3 r' \tens{G} \left( \vec{r}, \vec{r}', \omega \right) \sprod \hat{\vec{j}}_{\textrm{N}} \left( \vec{r}', \omega \right)}
\label{eq:Definition Electric Field}
\end{equation}
combines the Green's tensor with the properties stated in Eqs.~\eqref{eq:Lorentz Reciprocity} and \eqref{eq:Definition Real Part Imaginary Part}. The Green's tensor fulfills the relation $\tens{G} \left( \vec{r}, \vec{r}', \omega \right) \rightarrow \tens{0}$ for $\left| \vec{r} - \vec{r}' \right| \rightarrow \infty$ and the Schwarz reflection principle
\begin{equation}
\tens{G}^* \left( \vec{r}, \vec{r}', \omega \right) = \tens{G} \left( \vec{r}, \vec{r}', -\omega^* \right) \;\; \forall \; \vec{r}, \vec{r}', \omega.
\label{eq:Schwarz Reflection Principle}
\end{equation}
The respective equation for the magnetic field reads according to Eq.~\eqref{eq:Maxwell Equations}
\begin{equation}
\hat{\vec{B}} \left( \vec{r}, \omega \right) = \mu_0 \overrightarrow{\nabla} \vprod \int{\dif^3 r' \tens{G} \left( \vec{r}, \vec{r}', \omega \right) \sprod \hat{\vec{j}}_{\textrm{N}} \left( \vec{r}', \omega \right)}.
\label{eq:Definition Magnetic Field}
\end{equation}
The conductivity tensor $\tens{Q}$ from Eq.~\eqref{eq:Ohm's Law} and the Green's tensor from Eq.~\eqref{eq:Definition Electric Field} are related by
\begin{multline}
\mu_0 \omega \int \dif^3 s \int \dif^3 s' \tens{G} \left( \vec{r}, \vec{s}, \omega \right) \sprod \Re \left[ \tens{Q} \left( \vec{s}, \vec{s}', \omega \right) \right] \sprod \tens{G}^{*T} \left( \vec{r}', \vec{s}', \omega \right)\\
= \Im \left[ \tens{G} \left( \vec{r}, \vec{r}', \omega \right) \right],
\label{eq:Conductivity and Greens Tensor}
\end{multline}
where the definitions of the real and imaginary parts \eqref{eq:Definition Real Part Imaginary Part} are applied. The noise current $\hat{\vec{j}}_{\textrm{N}}$ from Ohm's law \eqref{eq:Ohm's Law} and the expression of the electric field \eqref{eq:Definition Electric Field} is defined as
\begin{equation}
\hat{\vec{j}}_{\textrm{N}} \left( \vec{r}, \omega \right) = \sqrt{\frac{\hbar \omega}{\pi}} \int{\dif^3 r' \tens{R} \left( \vec{r}, \vec{r}', \omega \right) \sprod \hat{\vec{f}} \left( \vec{r}', \omega \right)}
\label{eq:Noise Current}
\end{equation}
with the connection between the conductivity matrix $\tens{Q}$ and the $\tens{R}$-tensor given by
\begin{equation}
\int \dif^3 r' \tens{R} \left( \vec{r}, \vec{r}', \omega \right) \sprod \tens{R}^{*T} \left( \vec{r}'', \vec{r}', \omega \right) = \Re \left[ \tens{Q} \left( \vec{r}, \vec{r}'', \omega \right) \right].
\label{eq:Conductivity and R Matrix}
\end{equation}
The definition in Eq.~\eqref{eq:Conductivity and R Matrix} ensures that the fluctuation--dissipation theorem is obeyed. The noise current $\hat{\vec{j}}_{\textrm{N}}$ \eqref{eq:Noise Current} contains the creation and annihilation operators $\hat{\vec{f}}^{\dagger}$ and $\hat{\vec{f}}$, which satisfy the commutation relation
\begin{equation}
\left[ \hat{\vec{f}} \left( \vec{r}, \omega \right), \hat{\vec{f}}^{\dagger} \left( \vec{r}', \omega' \right) \right] = \textrm{\Greektens{$\delta$}} \left( \vec{r} - \vec{r}' \right) \delta \left( \omega - \omega' \right).
\end{equation}
The ground state expectation value of the creation and annihilation operators is given by
\begin{equation}
\langle \hat{\vec{f}} \left( \vec{r}, \omega \right) \hat{\vec{f}}^{\dagger} \left( \vec{r}', \omega' \right) \rangle = \textrm{\Greektens{$\delta$}} \left( \vec{r} - \vec{r}' \right) \delta \left( \omega - \omega' \right).
\label{eq:Expectation Values Operators}
\end{equation}
Eq.~\eqref{eq:Expectation Values Operators} will be used to derive an expression for the Casimir force for nonreciprocal material in Sec.~\ref{sec:Casimir Force for Nonreciprocal Media}. Afterwards we apply this general formula to the specific geometry of two semi-infinite slabs and derive the corresponding Green's tensor as solution of the Helmholtz equation \eqref{eq:Helmholtz Equation} with Eq.~\eqref{eq:Definition Electric Field} in Sec.~\ref{sec:Greens Tensor Two Surfaces}.

\section{Casimir Force for Nonreciprocal Media}
\label{sec:Casimir Force for Nonreciprocal Media}
Based on the results from Secs.~\ref{sec:Macroscopic QED Nonreciprocal Media} the Casimir force can be derived by using the Lorentz force acting on the internal charge $\hat{\rho}_{\textrm{in}}$ and current densities $\hat{\vec{j}}_{\textrm{in}}$ of a body
\begin{equation}
\hat{\vec{F}} = \int \limits_V {\dif^3 r \left( \hat{\rho}_{\textrm{in}} \hat{\vec{E}} + \hat{\vec{j}}_{\textrm{in}} \vprod \hat{\vec{B}} \right)}.
\end{equation}
By using Maxwell's equations \eqref{eq:Maxwell Equations}, these quantities can be expressed in terms of the electric and magnetic fields $\hat{\vec{E}}$ \eqref{eq:Definition Electric Field} and $\hat{\vec{B}}$ \eqref{eq:Definition Magnetic Field}, where the frequency components of the electric field $\hat{\vec{E}} \left( \vec{r}, \omega \right)$ and the total field $\hat{\vec{E}} \left( \vec{r} \right)$ are connected by the expression
\begin{equation}
\hat{\vec{E}} \left( \vec{r} \right) = \int \limits^{\infty}_0 \dif \omega \left[ \hat{\vec{E}} \left( \vec{r}, \omega \right) + \hat{\vec{E}}^{\dagger} \left( \vec{r}, \omega \right) \right].
\end{equation}
A similar expression also holds for the magnetic field components $\hat{\vec{B}} \left( \vec{r}, \omega \right)$. The Casimir force is the ground-state expectation value of the Lorentz force \cite{Buhmann_Book_1}. We apply the relation for vectors such as the electric and magnetic fields, exemplified for the electric field as $\overrightarrow{\nabla} \vec{E}^2 = 2 \left( \vec{E} \sprod \overrightarrow{\nabla} \right) \vec{E} + 2 \vec{E} \vprod \left( \overrightarrow{\nabla} \vprod \vec{E} \right)$ and note that the time-derivative of the ground-state average of the term $\hat{\vec{E}} \left( \vec{r} \right) \vprod \hat{\vec{B}} \left( \vec{r} \right)$ vanishes. Thus we obtain an expression for the Casimir force
\begin{multline}
\vec{F} = \int \limits_{\partial V} \dif \vec{A} \sprod \langle \epsilon_0 \hat{\vec{E}} \left( \vec{r} \right) \hat{\vec{E}} \left( \vec{r}' \right) + \frac{1}{\mu_0} \hat{\vec{B}} \left( \vec{r} \right) \hat{\vec{B}} \left( \vec{r}' \right)\\
- \frac{1}{2} \left[ \epsilon_0 \hat{\vec{E}} \left( \vec{r} \right) \sprod \hat{\vec{E}} \left( \vec{r}' \right) - \frac{1}{\mu_0} \hat{\vec{B}} \left( \vec{r} \right) \sprod \hat{\vec{B}} \left( \vec{r}' \right) \right] \tens{1} \rangle_{\vec{r}' \rightarrow \vec{r}},
\label{eq:Casimir Force 1}
\end{multline}
where $\tens{1}$ represents the unitary matrix. The transition $\vec{r}' \rightarrow \vec{r}$ is necessary because we must not include self-forces. Equation \eqref{eq:Casimir Force 1} holds both for reciprocal and nonreciprocal material and serves as a starting point for the following calculations.\\
We compute the expectation values of the electric and magnetic field components by making use of Eqs.~\eqref{eq:Definition Electric Field}, \eqref{eq:Definition Magnetic Field}, \eqref{eq:Conductivity and Greens Tensor}, \eqref{eq:Noise Current}, \eqref{eq:Conductivity and R Matrix} and use the results for the expectation values of the creation and annihilation operators \eqref{eq:Expectation Values Operators}
\begin{align}
\begin{array}{lll}
&\langle \hat{\vec{E}} \left( \vec{r} \right) \hat{\vec{E}} \left( \vec{r}' \right) \rangle &= \int \limits^{\infty}_0{\dif \omega \frac{\hbar \mu_0 \omega^2}{\pi} \Im \left[ \tens{G} \left( \vec{r}, \vec{r}', \omega \right) \right]}\\
&\langle \hat{\vec{B}} \left( \vec{r} \right) \hat{\vec{B}} \left( \vec{r}' \right) \rangle &= - \int\limits^{\infty}_0{\dif \omega \frac{\hbar \mu_0}{\pi} \overrightarrow{\nabla} \vprod \Im \left[ \tens{G} \left( \vec{r}, \vec{r}', \omega \right) \right] \vprod \overleftarrow{\nabla}'}.
\end{array}
\end{align}
The frequency integrals over the Green's tensor expressions can be evaluated further in the complex plane. By using the Schwarz reflection principle \eqref{eq:Schwarz Reflection Principle} and neglecting the values of the Green's tensor for large frequencies, $\lim_{|\omega| \rightarrow \infty} \frac{\omega^2}{c^2} \tens{G} = 0$, the expression for the Casimir force \eqref{eq:Casimir Force 1} at the frequency $\omega = \mi \xi$ yields 
\begin{multline}
\vec{F} = -\frac{\hbar}{2 \pi} \int\limits^{\infty}_0 \dif \xi \int \limits_{\partial V} \dif \vec{A} \cdot \left\{ \frac{\xi^2}{c^2} \tens{G}^{(1)} \left( \vec{r}, \vec{r}', \mi \xi \right) \right.\\
+ \frac{\xi^2}{c^2} \tens{G}^{(1) \textrm{T}} \left( \vec{r}', \vec{r}, \mi \xi \right)\\
+ \overrightarrow{\nabla} \times \tens{G}^{(1)} \left( \vec{r}, \vec{r}', \mi \xi \right) \times \overleftarrow{\nabla}' + \overrightarrow{\nabla} \times \tens{G}^{(1) \textrm{T}} \left( \vec{r}', \vec{r}, \mi \xi \right) \times \overleftarrow{\nabla}'\\
\left. -\textrm{Tr} \left[ \frac{\xi^2}{c^2} \tens{G}^{(1)} \left( \vec{r}, \vec{r}', \mi \xi \right) + \overrightarrow{\nabla} \times \tens{G}^{(1)} \left( \vec{r}, \vec{r}', \mi \xi \right) \times \overleftarrow{\nabla}' \right] \tens{1} \right\}_{\vec{r}' \rightarrow \vec{r}}.
\label{eq:Casimir Force 2}
\end{multline}
This is the first main result of this paper because Eq.~\eqref{eq:Casimir Force 2} generalizes the expression for the Casimir force to nonreciprocal media and holds for arbitrary geometrical properties. It differs from the respective result for reciprocal material by the presence of the transposed Green's tensor. In case of reciprocal material, Lorentz's reciprocity \eqref{eq:Lorentz Reciprocity} holds and there is no need for using the transposed version of the Green's tensor.\\
In Sec.~\ref{sec:Greens Tensor Two Surfaces}, the Green's tensor for the specific geometry of two semi-infinite plates, which are isotropic on the surface, is derived. Afterwards Eq.~\eqref{eq:Casimir Force 2} is applied to this specific geometry.

\section{Green's Tensor and Casimir Force for Two Planar Surfaces}
\label{sec:Greens Tensor Two Surfaces}
Having derived the general equations for the extension of the Casimir force to nonreciprocal materials, the scattering part of the Green's tensor for a setup consisting of two infinitely extended slabs separated by a distance $L$ is analyzed in this Section. The scattering part of the Green's tensor $\tens{G}^{(1)}$ of one planar surface is the integral over $\vec{k}^{\parallel}$ and contains the sum over the polarizations $\sigma$ of the incoming plane waves
\begin{equation}
\vec{a}_{k \pm \sigma} = \vec{e}_{\sigma \pm} \me^{\mi \left( \vec{k}^{\parallel} \sprod \vec{r} \pm k^{\perp} z \right)}
\end{equation}
and the respective sum over the polarizations $\sigma'$ of the outgoing wave
\begin{equation}
\vec{c}_{k \pm \sigma'} = \frac{\mi}{8 \pi^2 k^{\perp}} \vec{e}_{\sigma' \pm} \me^{-\mi \left( \vec{k}^{\parallel} \sprod \vec{r}' \pm k^{\perp} z' \right)}.
\end{equation}
Here we split the total wave vector $\vec{k}$ into its parallel component $\vec{k}^{\parallel}$, consisting of its $x$- ($k_x$) and $y$-components ($k_y$), and its $z$-component $k^{\perp}$. The unitary vectors of perpendicular polarization $\vec{e}_{\textrm{s} \pm}$ and parallel polarization $\vec{e}_{\textrm{p} \pm}$ are expressed in terms of the part of the wave vector parallel to the interface $\left( \vec{e}_{k^{\parallel}} \right)$ and perpendicular to it $\left( \vec{e}_z \right)$
\begin{align}
\begin{array}{lll}
&\vec{e}_{\textrm{s} \pm} &= \vec{e}_{k^{\parallel}} \vprod \vec{e}_z = \frac{1}{k^{\parallel}} \begin{pmatrix} k_y \\ -k_x \\ 0 \end{pmatrix}\\[6mm]
&\vec{e}_{\textrm{p} \pm} &= \frac{1}{k} \left( k^{\parallel} \vec{e}_z \mp k^{\perp} \vec{e}_{k^{\parallel}} \right) = \frac{1}{k} \begin{pmatrix} \mp \frac{k^{\perp}}{k^{\parallel}} k_x \\ \mp \frac{k^{\perp}}{k^{\parallel}} k_y \\ k^{\parallel} \end{pmatrix}.
\label{eq:Unitary Vectors}
\end{array}
\end{align}
The indices $\pm$ in the unitary vectors in Eq.~\eqref{eq:Unitary Vectors} refer to the directions of incoming and outgoing waves. In principle we can distinguish between four different possibilities which contribute to the final expression of the Green's tensor: odd/even number of reflections and an outgoing wave going to the left/right. A reflection coefficient $r^{\pm}_{\sigma, \sigma'}$ is added for each reflection at the left/right boundary, where the polarizability can be switched. The index $+ \left( - \right)$ refers to a reflection at the right (left) boundary. All combinations contribute to the final expression of the Green's tensor.\\
\begin{figure}[!ht]
\centerline{\includegraphics[width=\columnwidth]{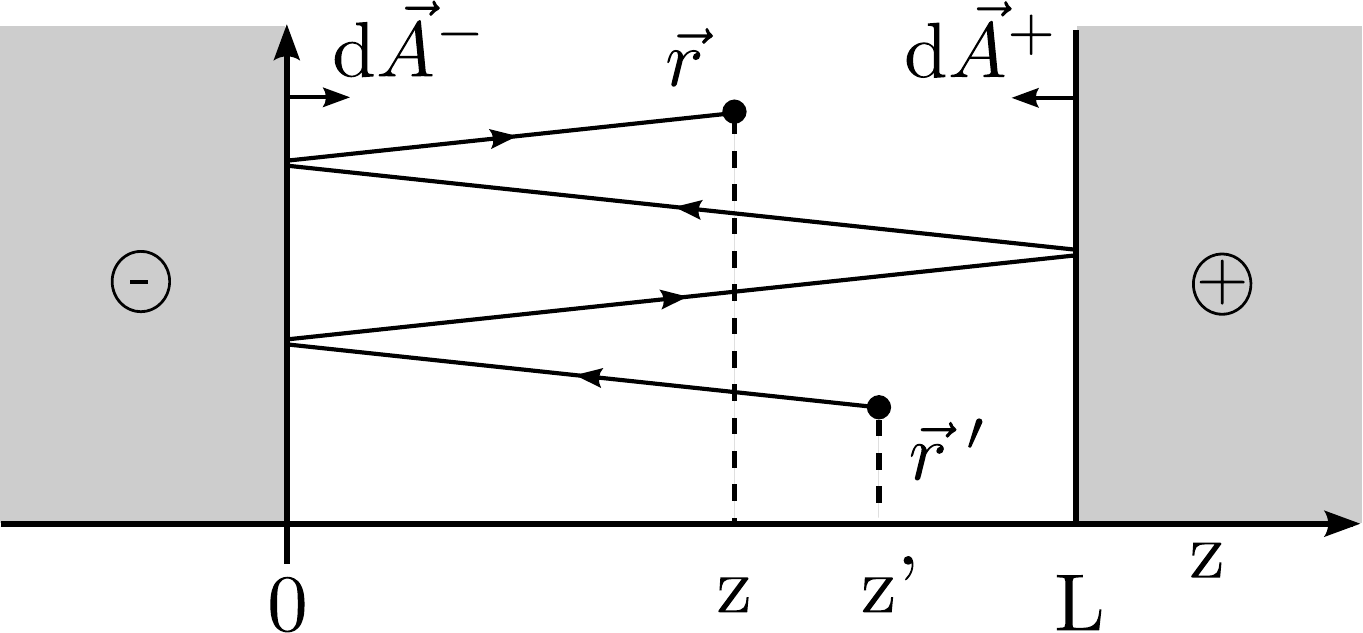}}
\caption{Sketch of the two semi-infinite half spaces with their boundaries at $z=0$ and $z=L$. The left half space is denoted by a minus sign and the right one by a plus sign. Moreover, the vectors $\dif \vec{A}^-$ and $\dif \vec{A}^+$ are orthogonal to the interfaces. The source point is located at $\vec{r}'$ and the field point at $\vec{r}$. There are three reflections in total. The outgoing wave goes in the left direction and the incoming one goes to the right.}
\label{fig:Figure1}
\end{figure}
The sketch in Fig.~\ref{fig:Figure1} shows three reflections, which appear in the Green's tensor beginning at the right. The Green's tensor is computed in the gap between the two semi-infinite plates with the boundaries at $z=0$ and $z=L$ and contains the unitary vectors $\vec{e}_{\sigma +} \vec{e}_{\sigma' -}$. The source of the wave is located at position $\vec{r}'$ and the field point is situated at $\vec{r}$, respectively. When the outgoing wave starts at the source at $z'$ and ends at $0$, the direction of the wave vector $k^{\perp}$ is negative leading to a contribution of $\me^{\mi k^{\perp} z'}$. A reflection with possible change of polarizability takes place at the left boundary. The subsequent path from $0$ to $L$ in the positive direction gives $\me^{\mi k^{\perp} L}$ and so does the following negative path from $L$ to $0$ after another reflection at the right boundary again. After the third reflection on the left side the final part from $0$ to $z$ is positive again giving rise to the factor $\me^{\mi k^{\perp} z}$.\\
The following expression contains the terms for one and three reflections
\begin{equation}
\me^{\mi k^{\perp} \left( z+z' \right)} r^-_{\sigma, \sigma'} + \sum \limits_{\sigma_1, \sigma_2} r^-_{\sigma, \sigma_1} r^+_{\sigma_1, \sigma_2} r^-_{\sigma_2, \sigma'} \me^{2 \mi k^{\perp} L} \me^{\mi k^{\perp} \left( z+z' \right)} + ...
\label{eq:Expression Reflection Coefficients}
\end{equation}
By introducing the reflection matrices for the right boundary $\mathcal{R}^+$ and the left boundary $\mathcal{R}^-$
\begin{equation}
\mathcal{R}^+ = \begin{pmatrix} r^+_{\textrm{s}, \textrm{s}} & r^+_{\textrm{s}, \textrm{p}} \\ r^+_{\textrm{p}, \textrm{s}} & r^+_{\textrm{p}, \textrm{p}} \end{pmatrix}; \;\; \mathcal{R}^- = \begin{pmatrix} r^-_{\textrm{s}, \textrm{s}} & r^-_{\textrm{s}, \textrm{p}} \\ r^-_{\textrm{p}, \textrm{s}} & r^-_{\textrm{p}, \textrm{p}}\end{pmatrix}
\label{eq:Reflection Matrix}
\end{equation}
Eq.~\eqref{eq:Expression Reflection Coefficients} for an infinite number of reflections can be rewritten as
\begin{equation}
\me^{\mi k^{\perp} \left( z+z' \right)} \left[ \mathcal{R}^- \sprod \sum \limits^{\infty}_{n=0} \left( \mathcal{R}^+ \sprod \mathcal{R}^- \me^{2 \mi k^{\perp} L} \right)^n \right]_{\sigma, \sigma'}.
\end{equation}
Analogous to the geometric sum for scalars, we define the infinite Neumann series under the assumption $|r^{\pm}_{\sigma, \sigma'}| \leq 1$ as
\begin{align}
\begin{array}{lll}
&\left( \mathcal{D}^+ \right)^{-1} &=\sum \limits^{\infty}_{n=0} \left( \mathcal{R}^+ \sprod \mathcal{R}^- \me^{2 \mi k^{\perp} L} \right)^n\\[2mm]
& &= \left[ 1 - \mathcal{R}^+ \sprod \mathcal{R}^- \me^{2 \mi k^{\perp} L} \right]^{-1}\\[4mm]
&\left( \mathcal{D}^- \right)^{-1} &=\sum \limits^{\infty}_{n=0} \left( \mathcal{R}^- \sprod \mathcal{R}^+ \me^{2 \mi k^{\perp} L} \right)^n\\[2mm]
& &= \left[ 1 - \mathcal{R}^- \sprod \mathcal{R}^+ \me^{2 \mi k^{\perp} L} \right]^{-1}.
\label{eq:Multiple Reflections}
\end{array}
\end{align}
After carrying out the same steps for the three other combinations, the Green's tensor eventually reads
\begin{multline}
\tens{G}^{(1)} \left( \vec{r}, \vec{r} \: ', \omega \right) = \frac{\mi}{8 \pi^2} \int \dif^2 k^{\parallel} \frac{1}{ k^{\perp}} \me^{\mi \vec{k}^{\parallel} \cdot \left( \vec{r}-\vec{r} \: ' \right)}\\[4mm]
\left[ \me^{\mi k^{\perp} \left( z+z' \right)} \begin{pmatrix} \vec{e}_{\textrm{s} +}, & \vec{e}_{\textrm{p} +} \end{pmatrix} \cdot \mathcal{R}^- \cdot \left( \mathcal{D}^+ \right)^{-1} \cdot \begin{pmatrix} \vec{e}_{\textrm{s}-} \\ \vec{e}_{\textrm{p}-} \end{pmatrix} \right.\\[4mm]
+ \me^{2 \mi k^{\perp} L} \me^{-\mi k^{\perp} \left( z+z' \right)} \begin{pmatrix} \vec{e}_{\textrm{s} -}, & \vec{e}_{\textrm{p} -} \end{pmatrix} \cdot \mathcal{R}^+ \cdot \left( \mathcal{D}^- \right)^{-1} \cdot \begin{pmatrix} \vec{e}_{\textrm{s} +} \\ \vec{e}_{\textrm{p} +} \end{pmatrix}\\[4mm]
+ \me^{2 \mi k^{\perp} L} \me^{\mi k^{\perp} \left( z-z' \right)} \begin{pmatrix} \vec{e}_{\textrm{s} +}, & \vec{e}_{\textrm{p} +} \end{pmatrix} \cdot \mathcal{R}^- \cdot \left( \mathcal{D}^+ \right)^{-1} \cdot \mathcal{R}^+ \cdot \begin{pmatrix} \vec{e}_{\textrm{s} +} \\ \vec{e}_{\textrm{p} +} \end{pmatrix}\\[4mm]
\left. + \me^{2 \mi k^{\perp} L} \me^{-\mi k^{\perp} \left( z-z' \right)} \begin{pmatrix} \vec{e}_{\textrm{s} -}, \vec{e}_{\textrm{p} -} \end{pmatrix} \cdot \mathcal{R}^+ \cdot \left( \mathcal{D}^- \right)^{-1} \cdot \mathcal{R}^- \cdot \begin{pmatrix}  \vec{e}_{\textrm{s} -} \\ \vec{e}_{\textrm{p} -} \end{pmatrix} \right].
\label{eq:Expression Greens Tensor}
\end{multline}
The scattering Green's tensor for a multi-layer system for reciprocal media was derived in Ref.~\cite{Tomas:1995}. Equation \eqref{eq:Expression Greens Tensor} is the most general expression of the scattering part of the Green's tensor for the setup shown in Fig.~\ref{fig:Figure1}. The first two terms in Eq.~\eqref{eq:Expression Greens Tensor} represent the contributions from an odd number of reflections containing one reflection coefficient apart from the infinite series. The last two terms are the contributions stemming from an even number of reflections showing two reflection coefficients apart from the series expression. Equation \eqref{eq:Expression Greens Tensor} contains several sums over polarizations allowing for a change of polarization at each boundary. The unitary vectors show the direction of the outgoing and incoming waves.\\
To express the Green's tensor \eqref{eq:Expression Greens Tensor} in Cartesian coordinates, the two-dimensional integral over the parallel component of the wave vector is transformed to
\begin{equation}
\int \dif^2 k^{\parallel} = \int \limits^{\infty}_0 \dif k^{\parallel} k^{\parallel} \int \limits^{2 \pi}_0\dif \phi.
\end{equation}
If the reflection coefficients show a specific symmetry, depend only on the absolute value of the parallel component of the wave vector $r_{\sigma, \sigma'} = r_{\sigma, \sigma'} \left( k^{\parallel} \right)$ and are independent of the angle $\phi$, then the angular contribution can be evaluated using the relations
\begin{align}
\begin{array}{lll}
\int\limits^{2 \pi}_0{\dif \phi \vec{e}_{\textrm{s} \pm}} \vec{e}_{\textrm{s} \pm}  &= \int\limits^{2 \pi}_0{\dif \phi \vec{e}_{\textrm{s} \pm}} \vec{e}_{\textrm{s} \mp} = \pi \left[ \vec{e}_x \vec{e}_x + \vec{e}_y \vec{e}_y \right]\\
\int\limits^{2 \pi}_0{\dif \phi \vec{e}_{\textrm{p} \pm} \vec{e}_{\textrm{p} \pm}} &= \frac{\pi}{k^2} \left[ \left( k^{\perp} \right)^2 \left( \vec{e}_x \vec{e}_x + \vec{e}_y \vec{e}_y \right) + 2 \left( k^{\parallel} \right)^2 \vec{e}_z \vec{e}_z \right]\\
\int\limits^{2 \pi}_0{\dif \phi \vec{e}_{\textrm{p} \pm} \vec{e}_{\textrm{p} \mp}} &= \frac{\pi}{k^2} \left[ - \left( k^{\perp} \right)^2 \left( \vec{e}_x \vec{e}_x + \vec{e}_y \vec{e}_y \right) + 2 \left( k^{\parallel} \right)^2 \vec{e}_z \vec{e}_z \right]\\
\int\limits^{2 \pi}_0{\dif \phi \vec{e}_{\textrm{s} \pm} \vec{e}_{\textrm{p} \pm}} &= \frac{\pi k^{\perp}}{k} \left[ \mp \vec{e}_x \vec{e}_y \pm \vec{e}_y \vec{e}_x \right]\\
\int\limits^{2 \pi}_0{\dif \phi \vec{e}_{\textrm{s} \pm} \vec{e}_{\textrm{p} \mp}} &= \frac{\pi k^{\perp}}{k} \left[ \pm \vec{e}_x \vec{e}_y \mp \vec{e}_y \vec{e}_x \right].
\label{eq:Unitary Vector Cartesian}
\end{array}
\end{align}
In Sec.~\ref{sec:Theoretical Model of the Photonic Topological Insulator}, the Casimir force is applied to a photonic topological insulator with a magnetic field pointing in the z-direction, for which Eq.~\eqref{eq:Unitary Vector Cartesian} and $r_{\sigma, \sigma'} = r_{\sigma, \sigma'} \left( k^{\parallel} \right)$ hold. For a magnetic field in the $xy$-plane, this expression would not hold anymore and one would need to use the more general expression in Eq.~\eqref{eq:Expression Greens Tensor}.\\
From these relations, it is apparent that the final expression of the Green's tensor shows diagonal contributions and a $xy$-component. There are no off-diagonal contributions involving the z-component, which is essential for the calculation of the Casimir force for the setup in Fig.~\ref{fig:Figure1}.\\
Equation \eqref{eq:Casimir Force 2} is evaluated for the case of two semi-infinite planes by inserting the expression for the Green's tensor \eqref{eq:Expression Greens Tensor}. In contrast to the respective result for reciprocal material \cite{Buhmann_Book_1} this expression includes the transposed Green's tensor. This is a consequence of the specific definition of the real and imaginary parts for nonreciprocal media \eqref{eq:Definition Real Part Imaginary Part}. Since only the diagonal elements of the Green’s tensor are non-zero due to the geometry consisting of two nonreciprocal planar infinite surfaces separated by vacuum, the expression of the Casimir force \eqref{eq:Casimir Force 2} does not differ in form from the respective reciprocal one. The nonreciprocity of the Casimir force arises from the specific expressions for the diagonal elements of the Green's tensor, containing the reflection matrices \eqref{eq:Reflection Matrix}.\\
To compute the contribution of the Casimir force \eqref{eq:Casimir Force 2} stemming from the magnetic field \eqref{eq:Definition Magnetic Field}, one has to apply the operators $\overrightarrow{\nabla}$ and $\overleftarrow{\nabla}$ to the unitary vectors \eqref{eq:Unitary Vectors} from the left and the right, respectively, with $\overrightarrow{\nabla} \vprod \rightarrow \mi \vec{k}_{\pm}$ and $\vprod \overleftarrow{\nabla} \rightarrow - \mi \vec{k}_{\pm}$. The unitary vectors are related as following
\begin{equation}
\mi \vec{k}_{\pm} \vprod \vec{e}_{\textrm{s} \pm} = \frac{\xi}{c} \vec{e}_{\textrm{p} \pm}, \; \; \mi \vec{k}_{\pm} \vprod \vec{e}_{\textrm{p} \pm} = -\frac{\xi}{c} \vec{e}_{\textrm{s} \pm}.
\end{equation}
After the substitution $k^{\perp} = \mi \kappa^{\perp}$ we finally obtain
\begin{multline}
\overrightarrow{\nabla} \times \tens{G}^{(1)} \left( \vec{r}, \vec{r} \: ', \mi \xi \right) \times \overleftarrow{\nabla}' = \frac{\xi^2}{8 \pi^2 c^2} \int{\dif^2 k^{\parallel} \frac{1}{\kappa^{\perp}} \me^{\mi \vec{k}^{\parallel} \cdot \left( \vec{r}-\vec{r} \: ' \right)}}\\[4mm]
\times \left[ \me^{- \kappa^{\perp} \left( z+z' \right)} \begin{pmatrix} \vec{e}_{\textrm{p} +}, & - \vec{e}_{\textrm{s} +} \end{pmatrix} \cdot \mathcal{R}^- \cdot \left( \mathcal{D}^+ \right)^{-1} \cdot \begin{pmatrix} \vec{e}_{\textrm{p} -} \\ - \vec{e}_{\textrm{s} -} \end{pmatrix} \right.\\[4mm]
+ \me^{- 2 \kappa^{\perp} L} \me^{\kappa^{\perp} \left( z + z' \right)} \begin{pmatrix} \vec{e}_{\textrm{p} -}, & - \vec{e}_{\textrm{s} -} \end{pmatrix} \cdot \mathcal{R}^+ \cdot \left( \mathcal{D}^- \right)^{-1} \cdot \begin{pmatrix} \vec{e}_{\textrm{p} +} \\ -\vec{e}_{\textrm{s} +} \end{pmatrix}\\[4mm]
+ \me^{- 2 \kappa^{\perp} L} \me^{- \kappa^{\perp} \left( z-z' \right)} \begin{pmatrix} \vec{e}_{\textrm{p} +},& -\vec{e}_{\textrm{s} +} \end{pmatrix}  \cdot \mathcal{R}^- \cdot \left( \mathcal{D}^+ \right)^{-1} \cdot \mathcal{R}^+ \cdot \begin{pmatrix} \vec{e}_{\textrm{p} +} \\ -\vec{e}_{\textrm{s} +} \end{pmatrix}\\[4mm]
\left. + \me^{- 2 \kappa^{\perp} L} \me^{\kappa^{\perp} \left( z - z' \right)} \begin{pmatrix} \vec{e}_{\textrm{p} -}, -\vec{e}_{\textrm{s} -} \end{pmatrix} \cdot \mathcal{R}^+ \cdot \left( \mathcal{D}^- \right)^{-1} \cdot \mathcal{R}^- \cdot \begin{pmatrix} \vec{e}_{\textrm{p} -} \\ -\vec{e}_{\textrm{s} -} \end{pmatrix} \right].
\end{multline}
The Casimir force for the setup in Fig.~\ref{fig:Figure1} is the force on the right plate situated at $z=L$. Due to our geometry the surface vector of the right plane points ourwards $\dif \vec{A} = - \dif A \vec{e}_z$. Since the surface has infinite extensions in the x- and y-directions, the total Casimir force $\vec{F}$ diverges. Thus we restrict ourselves to the calculation of the surface force density $\vec{f}$, which is equivalent to the Casimir pressure. Due to this geometry, the resulting term of the Casimir force only contains contributions in the z-direction. Since the z-contributions of the Green's tensor \eqref{eq:Expression Greens Tensor} do not show any mixing terms with the x- or y-components \eqref{eq:Unitary Vector Cartesian}, the zz-component of the Casimir force is the only relevant one.\\
Since the term of the Casimir force \eqref{eq:Casimir Force 2} shows the trace over the Green's tensor, one has to use the xx- and yy-components of the Green's tensor \eqref{eq:Expression Greens Tensor} beside its zz-component. After making use of the relation $\xi^2 = \left( \kappa^{\perp} \right)^2 - \left( k^{\parallel} \right)^2$, the Casimir force per unit area eventually reads
\begin{multline}
\vec{f} = -\frac{\hbar}{4 \pi^2} \int \limits^{\infty}_0 \dif \xi \int \limits^{\infty}_0 \dif k^{\parallel} k^{\parallel} \kappa^{\perp} e^{-2 \kappa^{\perp} L}\\
\times \mathrm{Tr} \left[ \mathcal{R}^- \cdot \left( \mathcal{D}^+ \right)^{-1} \cdot \mathcal{R}^+ + \mathcal{R}^+ \cdot \left( \mathcal{D}^- \right)^{-1} \cdot \mathcal{R}^- \right] \vec{e}_z.
\label{eq:Casimir Force Final Result}
\end{multline}
Since $\vec{f}$ always points in the z-direction, it is convenient to work with the scalar Casimir force $f$ defined by $\vec{f} = f \vec{e}_z$. It contains only terms which have their origins in an even number of reflections at the boundaries. Besides, the polarizations of the outgoing and incoming wave are the same, which does not necessarily mean that there is no polarization change between the first and the last reflection at the boundaries.\\
Carrying out both sums over polarizations one can find a more explicit form of Eq.~\eqref{eq:Casimir Force Final Result}
\begin{multline}
\vec{f} = -\frac{\hbar}{2 \pi^2} \int \limits^{\infty}_0 \dif \xi \int \limits^{\infty}_0 \dif k^{\parallel} k^{\parallel} \kappa^{\perp}\\
\times \me^{-2 \kappa^{\perp} L} \frac{a- 2\me^{-2 \kappa^{\perp} L}b }{1 - \me^{-2 \kappa^{\perp} L}\, a + \me^{-4 \kappa^{\perp} L} \, b } \vec{e}_z
\label{eq:Casimir Force Final Result Alternative}
\end{multline}
with
\begin{align}
\begin{array}{lll}
&a &= r^-_{\textrm{s}, \textrm{s}} r^+_{\textrm{s}, \textrm{s}} +r^-_{\textrm{p}, \textrm{p}} r^+_{\textrm{p}, \textrm{p}}+ r^-_{\textrm{p}, \textrm{s}} r^+_{\textrm{s}, \textrm{p}}+ r^-_{\textrm{s}, \textrm{p}} r^+_{\textrm{p}, \textrm{s}}\\[2mm]
&b &= \left( r^+_{\textrm{s}, \textrm{p}} r^+_{\textrm{p}, \textrm{s}} + r^+_{\textrm{s}, \textrm{s}} r^+_{\textrm{p}, \textrm{p}} \right) \left( r^-_{\textrm{s}, \textrm{p}} r^-_{\textrm{p}, \textrm{s}} + r^-_{\textrm{s}, \textrm{s}} r^-_{\textrm{p}, \textrm{p}} \right).
\end{array}
\end{align}
In the following, this general expression of the Casimir force for nonreciprocal material \eqref{eq:Casimir Force Final Result} or \eqref{eq:Casimir Force Final Result Alternative} is evaluated for the case of the photonic topological insulator InSb. Besides, it can also be used to reproduce the well-known results for the Casimir force for two perfectly conducting reciprocal mirrors with reflection coefficients $r^{\pm}_{\textrm{s}, \textrm{s}} = -1$, $r^{\pm}_{\textrm{p}, \textrm{p}} = +1$ and $r^{\pm}_{\textrm{p}, \textrm{s}} = r^{\pm}_{\textrm{s}, \textrm{p}} = 0$ and two perfectly reflecting nonreciprocal mirrors with $r^{\pm}_{\textrm{s}, \textrm{p}} = r^{\pm}_{\textrm{p}, \textrm{s}} = -1$ and $r^{\pm}_{\textrm{s}, \textrm{s}} = r^{\pm}_{\textrm{p}, \textrm{p}} = 0$. In both cases we obtain the final result
\begin{equation} \label{eq:Casimir Force perfect conduct}
\vec{f} = - \frac{\pi^2 \hbar c}{240 \, L^4} \vec{e}_z.
\end{equation}
The Casimir force between one perfectly conducting and one perfectly permeable plate with the reflection coefficients $r^-_{\textrm{s,s}} = r^-_{\textrm{p}, \textrm{p}} = -1$, $r^-_{\textrm{p}, \textrm{p}} = r^-_{\textrm{s}, \textrm{s}} = 1$ and $r^\pm_{\textrm{s}, \textrm{p}} = r^\pm_{\textrm{p}, \textrm{s}} = 0$ reads
\begin{equation}
\label{eq:Casimir Force perfect neg}
\vec{f} = \frac{7\pi^2 \hbar c}{1920\, L^4} \vec{e}_z
\end{equation}
and is repulsive, cf. Ref.~\cite{Boyer:1974}. The sign of the Casimir force thus depends on the material used.\\
The central result \eqref{eq:Casimir Force Final Result} can be applied to all kinds of nonreciprocal materials as long as their reflection coefficients are independent of the angle $\phi$. Both axion topological insulators, cf. Refs.~\cite{Wilczek:1987, Hasan:2010}, and photonic topological insulators with perpendicular bias fulfill this property. The Casimir force between axion topological insulators with frequency-dependent permittivity and permeability and frequency-independent axion coupling was studied in Ref.~\cite{Grushin:2011}, leading to the prediction of repulsive Casimir forces. A later treatment with a more involved material model \cite{Rodriguez-Lopez:2014} incorporates a frequency dependent axion coupling.\\
In Sec.~\ref{sec:Results}, Eq.~\eqref{eq:Casimir Force Final Result} is applied to a photonic topological insulator (PTI), whose material model is outlined in Sec.~\ref{sec:Theoretical Model of the Photonic Topological Insulator}.\\

\section{Theoretical Model of the Photonic Topological Insulator}
\label{sec:Theoretical Model of the Photonic Topological Insulator}
A photonic topological insulator (PTI) shows a mixing of polarizations, cf. Eq.~\eqref{eq:Reflection Matrix}, which stems from the PTI's anisotropic permittivity $\Greektens{$\epsilon$}$, cf. Sec.~\ref{sec:Antisymmetric Permittivity}. A specific material model for the permittivity is based on InSb and is explained in Sec.~\ref{sec:Material model for InSb}. Afterwards, the reflection coefficients \eqref{eq:Reflection Matrix} needed for the calculation of the Casimir force are derived in Sec.~\ref{sec:Reflection coefficients}.

\subsection{Antisymmetric Permittivity}
\label{sec:Antisymmetric Permittivity}
The mixing of polarizations stems from the PTI's anisotropic permittivity $\Greektens{$\epsilon$}$, which has the form of an antisymmetric tensor
\begin{equation}
\Greektens{$\epsilon$} = \begin{pmatrix} \epsilon_{xx} & \epsilon_{xy} & 0 \\ - \epsilon_{xy} & \epsilon_{xx} & 0 \\ 0 & 0 & \epsilon_{zz} \end{pmatrix}.
\label{eq:Epsilon}
\end{equation}
We assume a constant unit permeability $\mu = 1$. Furthermore all magnetoelectric cross susceptibilities are zero $\left( \zeta = \xi= 0 \right)$. The permittivity tensor \eqref{eq:Epsilon} of such materials is antisymmetric and thus violates Lorentz's reciprocity principle \eqref{eq:Lorentz Reciprocity} and consequently time-reversal symmetry.\\
This model is studied in the Voigt configuration, where the surface is perpendicular to the bias magnetic field $\vec{B}$ \cite{Hanson:2016}. The normal vector of the interface is parallel to the $z$-axis. In this particular case we find that the system is symmetric in the $xy$-plane because its $\epsilon$-tensor \eqref{eq:Epsilon} is rotationally invariant around the z-axis for an arbitrary angle $\phi$
\begin{equation}
\tens{R}^{\textrm{T}} \cdot \Greektens{$\epsilon$} \cdot \tens{R} = \Greektens{$\epsilon$} \quad 
	\text{with} \quad \tens{R} = \begin{pmatrix} \cos \left( \phi \right) & \sin \left( \phi \right) & 0\\
	-\sin \left( \phi \right) & \cos \left( \phi \right) & 0\\
	0 & 0 & 1 \end{pmatrix}.
\end{equation}
A particular challenge is the fact that in a PTI neither solely perpendicularly ($\textrm{s}$) nor solely parallelly ($\textrm{p}$) polarized waves are solutions of Maxwell's equations. Thus a more general approach is needed to find the electric field in the PTI. Mathematically this procedure is similar to the one for biaxial, anisotropic magnetodielectics \cite{Rosa:2008_2} and is presented in Sec.~\ref{sec:Reflection coefficients}.

\subsection{Material model for the Permittivity based on InSb}
\label{sec:Material model for InSb}
We compute the Casimir force of the PTI for a specific material model, which is based on $n$-doped InSb with an external static magnetic field pointing in z-direction, $\textbf{B} = B \vec{e}_z$. As was already mentioned in Sec.~\ref{sec:Greens Tensor Two Surfaces}, the magnetic field pointing in the z-direction is an essential condition for using Eq.~\eqref{eq:Casimir Force Final Result}. This material has been investigated in Ref.~\cite{Palik:1976}, and more recently with a higher doping in Ref.~\cite{Law:2014}. It has been used to study the near-field heat transfer by various authors, cf. Refs.~\cite{Moncada:2015, Ben:2016, Zhu:2016}. The entries of the permittivity tensor \eqref{eq:Epsilon} for this specific model read
\begin{align}
\begin{array}{lll}
\epsilon_{xx} =	\epsilon_{\textrm{inf}} \left\{ \displaystyle{\frac{\omega_{\textrm{p}}^2 \left( \omega + \mi \gamma \right)}{\omega \left[ \omega_{\textrm{c}}^2- \left( \omega + \mi \gamma \right)^2 \right]}} + \displaystyle{\frac{\omega_{\textrm{L}}^2-\omega_{\textrm{T}}^2}{-\mi \Gamma \omega + \omega_{\textrm{T}}^2 - \omega^2}}+1 \right\}\\[6mm]
\epsilon_{zz} = \epsilon_{\textrm{inf}} \left\{ -\displaystyle{\frac{\omega_{\textrm{p}}^2}{\omega \left( \omega + \mi \gamma \right)}} + \displaystyle{\frac{ \omega_{\textrm{L}}^2 - \omega_{\textrm{T}}^2}{- \mi \Gamma \omega - \omega^2 + \omega_{\textrm{T}}^2}} + 1 \right\}\\[6mm]
\epsilon_{xy} = \displaystyle{\frac{ \mi \epsilon_{\textrm{inf}} \omega_{\textrm{c}} \omega_{\textrm{p}}^2}{\omega \left[\omega_{\textrm{c}}^2- \left( \omega + \mi \gamma \right)^2 \right]}}.
\label{eq:Epsilon InSb}
\end{array}
\end{align}
The plasma and cyclotron frequencies are given by
\begin{align}
\begin{array}{lll}
&\omega_{\textrm{p}} &=	\sqrt{\displaystyle{\frac{n q_{\textrm{e}}^2}{\epsilon_{\textrm{inf}} m^\star \epsilon_0}}}\\[4mm]
&\omega_{\textrm{c}} &= \displaystyle{\frac{B q_{\textrm{e}}}{m^\star}}
\label{eq:Epsilon InSb Frequencies}
\end{array}
\end{align}
where $q_e$ is the electron charge and $m^\star$ its reduced mass. $\Gamma$ represents the phonon damping constant and $\gamma$ is the free carrier damping constant. Throughout this paper we will use the following values for the material constants which have been measured in \cite{Palik:1976}: $\omega_{\textrm{L}} = 3.62 \cdot 10^{13}$ rad/s, $\omega_{\textrm{T}} = 3.39 \cdot 10^{13}$ rad/s, $\Gamma = 5.65 \cdot 10^{11}$ rad/s, $\gamma = 3.39 \cdot 10^{12}$ rad/s, $n = 1.07 \cdot 10^{17} \: \text{cm}^{-3}$, $m^\star = 0.022 \cdot m_{\textrm{e}}$ where $m_{\textrm{e}}$ is the electron mass. Additionally, Ref.~\cite{Palik:1976} used $\epsilon_{\textrm{inf}} = 15.7$, but since we are integrating over all frequencies equally we have to set $\epsilon_{\textrm{inf}} = 1$ to ensure convergence. Physically this means we are neglecting certain resonances of $\epsilon(\omega)$ and only take contributions from the ones at $\omega_{\textrm{T}}$ and $\omega_{\textrm{c}}$ into account. Since we are interested in the influence of the offdiagonal elements of $\Greektens{$\epsilon$}$ on the Casimir force this seems to be a reasonable simplification, because the other resonances are not contributing to $\epsilon_{xy}$.\\
Note that the cyclotron frequency $\omega_{\textrm{c}}$ is proportional to the external magnetic field, so by changing its direction from $+\vec{e}_z$ to $-\vec{e}_z$ one can change the sign of $\omega_{\textrm{c}}$. This results in a sign change of the offdiagonal elements of $\Greektens{$\epsilon$}$ as one can see in Eq.~\eqref{eq:Epsilon InSb} and this is equivalent to applying the time reversal operator.\\
The permittivity model for InSb \eqref{eq:Epsilon InSb} with Eq.~\eqref{eq:Epsilon InSb Frequencies} is closely related to the one for a single-component magnetic plasma biased with an external static magnetic field $\vec{B} = B_z \vec{e}_z$, which is examined in Refs.~\cite{Hanson:2016, Bittencourt_Book}. By setting the damping constants $\gamma$ and $\Gamma$ equal to zero and if $\omega_{\textrm{L}} = \omega_{\textrm{T}}$, Eq.~\eqref{eq:Epsilon InSb} reduces to the model of the magnetic plasma. One could consider different parameter ranges with this material, e.g. for $\omega_{\textrm{p}}$. This model often applies to gas plasmas, because phonon contributions are ignored, but it is considered to be the simplistic model of a free-carrier material subject to a bias field.

\subsection{Reflection coefficients}
\label{sec:Reflection coefficients}
We consider a single interface, where the half space $z < 0$ is vacuum and the half space $z > 0$ is a PTI. In Ref.~\cite{Rosa:2008_2} the field in the vacuum is described by general amplitudes $e$, where the indices $\textrm{i}$ and $\textrm{r}$ refer to the incoming and reflected waves and the indices $\textrm{s}$ and $\textrm{p}$ specify the polarization. We define a $\textrm{s}$-polarized wave by $\vec{E} \parallel \vec{e}_y$ and a $\textrm{p}$-polarized one by $\vec{B} \parallel \vec{e}_y$. Using Maxwell's equations in cgs-units
\begin{equation}
\overrightarrow{\nabla} \times \vec{E} = -\frac{1}{c} \frac{\partial}{\partial t} \vec{H}, \quad \overrightarrow{\nabla} \times \vec{H}  = \frac{1}{c} \frac{\partial}{\partial t}\left( \Greektens{$\epsilon$} \cdot \vec{E} \right)
\label{eq:Maxwell Equations cgs}
\end{equation}
and setting $\epsilon = 1$ in the vacuum the equations for a general incoming wave with $\vec{k} = (k^x, 0, k^{\perp}_{\textrm{i}})^{\textrm{T}}$ read
\begin{align}
\begin{array}{lll}
&\vec{E}_{\textrm{i}} &= e_{\textrm{s}, \textrm{i}} \vec{e}_y + e_{\textrm{p}, \textrm{i}} \frac{c}{\omega} \left( k^{\perp}_{\textrm{i}} \vec{e}_x - k^x  \vec{e}_z \right) \me^{ \mi \left( k^x x + k^{\perp}_{\textrm{i}} z - \omega t \right)}\\[2mm]
&\vec{H}_{\textrm{i}} &= e_{\textrm{p}, \textrm{i}} \vec{e}_y - e_{\textrm{s}, \textrm{i}} \frac{c}{\omega} \left( k^{\perp}_{\textrm{i}} \vec{e}_x - k^x \vec{e}_z \right) \me^{\mi \left( k^x x + k^{\perp}_{\textrm{i}} z -\omega t \right)}\\[2mm]
&\vec{E}_{\textrm{r}} &= e_{\textrm{s}, \textrm{r}} \vec{e}_y - e_{\textrm{p}, \textrm{r}} \frac{c}{\omega} \left( k^{\perp}_{\textrm{i}} \vec{e}_x + k^x  \vec{e}_z \right) \me^{\mi \left( k^x x - k^{\perp}_{\textrm{i}} z -\omega t \right)}\\[2mm]
&\vec{H}_{\textrm{r}} &= e_{\textrm{p}, \textrm{r}} \vec{e}_y + e_{\textrm{s}, \textrm{r}} \frac{c}{\omega} \left( k^{\perp}_{\textrm{i}} \vec{e}_x + k^x \vec{e}_z \right) \me^{\mi \left( k^x x - k^{\perp}_{\textrm{i}} z -\omega t \right)}.
\label{eq:vacfields}
\end{array}
\end{align}
Since our setup is $xy$-symmetric we assumed without loss of generality $\vec{k}^{\parallel} = k^x \vec{e}_x$ and we have used $k^{\perp}_{\textrm{i}} = -k^{\perp}_{\textrm{r}}$.\\
Due to the structure of the $\Greektens{$\epsilon$}$-tensor the $\textrm{s}$- and $\textrm{p}$-polarized contributions cannot be separated from each other any longer. Thus in a more general approach plane waves are assumed
\begin{equation}
\vec{E} = \begin{pmatrix}
e_x(z) \\ e_y(z) \\ e_z(z) \end{pmatrix} \me^{\mi \left( k^x x - \omega t \right)}, \quad \vec{H} = \begin{pmatrix} h_x(z) \\ h_y(z) \\ h_z(z) \end{pmatrix} \me^{\mi \left( k^x x- \omega t \right)}.
\end{equation}
$k^x$ is conserved across the interface. The $z$-components of Maxwell's equations \eqref{eq:Maxwell Equations cgs} read
\begin{equation}
h_z = \frac{c}{\omega} k^x e_y, \quad e_z = -\frac{c}{\omega \epsilon_{zz}} k^x h_y
\label{eq:zKomponenten im PTI}
\end{equation}
and can be inserted into the $x,y$-contributions. For these components we introduce the vector $\vec{u}$ with $u_1= e_x$, $u_2=e_y$, $u_3=h_x$ and $u_4=h_y$. By assuming the ansatz $u_j = u_j \left( 0 \right) \me^{\mi k^{\perp} z}$ for the single components with $k^{\perp}$ as the $z$-contribution of the wave vector in the PTI one obtains again from Eq.~\eqref{eq:Maxwell Equations cgs}
\begin{equation}
\tens{L} \sprod \vec{u} = - \frac{c}{\omega} k^{\perp} \vec{u}
\end{equation}
with
\begin{equation}
\tens{L} = \begin{pmatrix} 0 & 0 & 0 & -1+\frac{c^2}{\omega^2 \epsilon_{zz}} \left( k^x \right)^2 \\
   0 & 0 & 1 & 0 \\- \epsilon_{xy} & \epsilon_{xx} - \frac{c^2}{\omega^2} \left( k^x \right)^2 & 0 & 0 \\ - \epsilon_{xx} & -\epsilon_{xy} & 0 & 0 \end{pmatrix}.
\label{eq:L Matrix}
\end{equation}
To find nontrivial solutions one has to solve the equation $\textrm{det} \left( \tens{L} + \frac{\omega k^{\perp}}{c} \tens{1} \right) =0 $ leading to the dispersion relations 
\begin{equation}
k^{\perp \left( m \right)} = \pm \frac{\omega}{c} \frac{1}{\sqrt{2}} \sqrt{A + B \pm \sqrt{(A -B)^2 + 4C}}
\label{eq:Dispersion Relation}
\end{equation}
with
\begin{align}
\begin{array}{lll}
&A &= \epsilon_{xx} \left[ 1- \displaystyle{\frac{c^2}{\omega^2\epsilon_{zz}}} \left( k^x \right)^2 \right]\\
&B &= \epsilon_{xx} - \displaystyle{\frac{c^2}{\omega^2}} \left( k^x \right)^2\\
&C & = - \left[ 1 - \displaystyle{\frac{c^2}{\omega^2\epsilon_{zz}}} \left( k^x \right)^2 \right] \epsilon_{xy}^2
\end{array}
\end{align}
for the four mathematical solutions $m=1,2,3,4$, corresponding to the four possible combinations of signs in Eq.~\eqref{eq:Dispersion Relation}. Since solutions with $\textrm{Re} \left( k^{\perp \left( m \right)} \right) < 0 $ would result in waves propagating in negative $z$-direction we can neglect these solutions for the transmitted wave propagating in the positive $z$-direction. Let $k^{\perp (1)}$ and $k^{\perp (2)}$ be the two solutions with positive real parts and neglect the other two ones and we finally arrive at the expression for the transmitted components of $\vec{E}$ and $\vec{H}$ parallel to the surface
\begin{equation}
\label{TIfields}
\begin{pmatrix} \vec{E}_{\textrm{t}} \\ \vec{H}_{\textrm{t}} \end{pmatrix} = \me^{\mi \left( k^x x - \omega t \right)} \sum\limits_{m=1,2} \vec{u}^{(m)} \left( 0 \right) \me^{\mi k^{\perp (m)} z}.
\end{equation} 
According to the continuity relations, the parallel components of the electric and magnetic fields $\vec{E}$ and $\vec{H}$, i.e. the $x$,$y$-components, at the interface between vacuum and the topological insulator are continuous. Since this set of four equations depends on too many variables we start by expressing $e_y$, $h_x$ and $h_y$ in terms of $e_x$ using Eq.~\eqref{eq:L Matrix} 
\begin{align}
\begin{array}{llll}
&\alpha^{(m)} &\equiv \displaystyle{\frac{e_y^{(m)} \left( 0 \right)}{e_x^{(m)} \left( 0 \right)}} &= \displaystyle{\frac{L_{23} L_{31}}{\frac{c^2 \left( k^{\perp (m)} \right)^2}{\omega^2}- L_{23}L_{32}}}\\
&\beta^{(m)} &\equiv \displaystyle{\frac{h_x^{(m)} \left( 0 \right)}{e_x^{(m)} \left( 0 \right)}} &= -\displaystyle{\frac{\omega}{c k^{\perp (m)}}} L_{31} - \displaystyle{\frac{\omega}{c k^{\perp (m)}}} L_{32} \alpha^{(m)}\\
&\gamma^{(m)} &\equiv \displaystyle{\frac{h_y^{(m)} \left( 0 \right)}{e_x^{(m)} \left( 0 \right)}} &= - \displaystyle{\frac{\omega}{c k^{\perp (m)}}} L_{41} - \displaystyle{\frac{\omega}{c k^{\perp (m)}}} L_{31} \alpha^{(m)}.
\end{array}
\end{align}
These equations are inserted into the boundary conditions and one obtains
\begin{equation}
\label{eq:PTIEquationforamplitudes}
\underbrace{\begin{pmatrix} -1 & 0 & \alpha^{(1)} & \alpha^{(2)}\\ \frac{c k^{\perp}_{\textrm{i}}}{\omega} & 0 & -\beta^{(1)} & -\beta^{(2)} \\	0 & \frac{c k^{\perp}_{\textrm{i}}}{\omega} & 1 & 1 \\ 0 & -1 & \gamma^{(1)} & \gamma^{(2)}	\end{pmatrix}}_{\equiv \tens{M}} \sprod \begin{pmatrix} e_{\textrm{s}, \textrm{r}} \\ e_{\textrm{p}, \textrm{r}} \\ e^{(1)}_x \\ e^{(2)}_x \end{pmatrix} = \begin{pmatrix} e_{\textrm{s}, \textrm{i}} \\ \frac{c k^{\perp}_{\textrm{i}}}{\omega} e_{\textrm{s}, \textrm{i}} \\ \frac{c k^{\perp}_{\textrm{i}}}{\omega} e_{\textrm{p}, \textrm{i}} \\ e_{\textrm{p}, \textrm{i}} \end{pmatrix}.
\end{equation}
We now assume the incoming wave separately for $\textrm{s}$- ($e_{\textrm{p}, \textrm{i}}=0$, $e_{\textrm{s}, \textrm{i}} \neq 0$) or $\textrm{p}$-polarization ($e_{\textrm{s}, \textrm{i}}=0$, $e_{\textrm{p}, \textrm{i}} \neq 0$) and solve for the reflected amplitudes to finally obtain the reflection coefficients by help of Kramers rule  
\begin{align} \label{eq:reflectioncoefficients}
\begin{array}{lll}
&r_{\textrm{s}, \textrm{s}} &= \displaystyle{\frac{e_{\textrm{s}, \textrm{r}}}{e_{\textrm{s}, \textrm{i}}}} = \displaystyle{\frac{\textrm{det} \left( \tens{M}_1 \right)}{\textrm{det} \left( \tens{M} \right)}}\\[4mm]
&r_{\textrm{p}, \textrm{s}} &= \displaystyle{\frac{e_{\textrm{p}, \textrm{r}}}{e_{\textrm{s}, \textrm{i}}}} = \displaystyle{\frac{\textrm{det} \left( \tens{M}_2 \right)}{\textrm{det} \left( \tens{M} \right)}}\\[4mm]
&r_{\textrm{s}, \textrm{p}} &= \displaystyle{\frac{e_{\textrm{s}, \textrm{r}}}{e_{\textrm{p}, \textrm{i}}}} = \displaystyle{\frac{\textrm{det} \left( \tens{M}_3 \right)}{\textrm{det} \left( \tens{M} \right)}}\\[4mm]
&r_{\textrm{s}, \textrm{s}} &= \displaystyle{\frac{e_{\textrm{p}, \textrm{r}}}{e_{\textrm{p}, \textrm{i}}}} = \displaystyle{\frac{\textrm{det} \left( \tens{M}_4 \right)}{\textrm{det} \left( \tens{M} \right)}}
\end{array}
\end{align}
with the matrices
\begin{align}
\begin{array}{lll}
&\tens{M}_1 &= \begin{pmatrix} 1 & 0 & \alpha^{(1)} & \alpha^{(2)} \\ \frac{c k^{\perp}_{\textrm{i}}}{\omega} & 0 & -\beta^{(1)} & -\beta^{(2)} \\ 0 & \frac{c k^{\perp}_{\textrm{i}}}{\omega} & 1 & 1 \\ 0 & -1 & \gamma^{(1)} & \gamma^{(2)} \end{pmatrix}\\[10mm]
&\tens{M}_2 &= \begin{pmatrix} -1 & 1 & \alpha^{(1)} & \alpha^{(2)} \\ \frac{c k^{\perp}_{\textrm{i}}}{\omega} & \frac{c k^{\perp}_{\textrm{i}}}{\omega}& -\beta^{(1)} & -\beta^{(2)} \\ 0 & 0 & 1 & 1 \\ 0 & 0 & \gamma^{(1)} & \gamma^{(2)} \end{pmatrix}\\[10mm]
&\tens{M}_3 &= \begin{pmatrix} 0 & 0 & \alpha^{(1)} & \alpha^{(2)} \\ 0 & 0 & -\beta^{(1)} & -\beta^{(2)} \\ \frac{c k^{\perp}_{\textrm{i}}}{\omega} & \frac{c k^{\perp}_{\textrm{i}}}{\omega} & 1 & 1 \\ 1 & -1 & \gamma^{(1)} & \gamma^{(2)} \end{pmatrix}\\[10mm]
&\tens{M}_4 &= \begin{pmatrix} -1 & 0 & \alpha^{(1)} & \alpha^{(2)} \\ \frac{c k^{\perp}_{\textrm{i}}}{\omega} & 0 & -\beta^{(1)} & -\beta^{(2)} \\ 0 & \frac{c k^{\perp}_{\textrm{i}}}{\omega} & 1 & 1 \\ 0 & 1 & \gamma^{(1)} & \gamma^{(2)} \end{pmatrix}.
\end{array}
\end{align}
Since det$\left( \tens{M}_2 \right)= $ det$\left( \tens{M}_3 \right)$, the off-diagonal reflection coefficients are equal, $r_{\textrm{s}, \textrm{p}} = r_{\textrm{p}, \textrm{s}}$. Moreover by changing the sign of $\epsilon_{xy}$, which is equivalent to applying the time-reversal operator, we obtain: $r_{\sigma, \bar{\sigma}} \to - r_{\sigma, \bar{\sigma}}$ and  $r_{\sigma, \sigma} \to  r_{\sigma, \sigma}$ with $\sigma = \textrm{s, p}$ and $\bar{\sigma} = \textrm{p,s}$, respectively. By setting $\epsilon_{xy} = 0 $ we find that the reflection coefficients simplify to the case of a uniaxial out of plane metamaterial \cite{Rosa:2008_2}, especially if $r_{\textrm{p}, \textrm{s}} = r_{\textrm{s}, \textrm{p}} = 0$ and the nonreciprocity vanishes.
By assuming the model described by Eq.~\eqref{eq:Epsilon InSb} for InSb we can change the sign of  $r_{\sigma, \bar{\sigma}}$ simply by changing the sign of the magnetic field and moreover, by switching the magnetic field on and off we can switch between a reciprocal and a nonreciprocal case.\\
Furthermore we consider the case where we exchange the positions of the TI and the vacuum which is equivalent to changing the $z$ coordinate to $-z$. We still consider the incoming wave to propagate in the vacuum but this time in the $-z$ direction before it is reflected by the TI. So in this case we have to exchange $k^{\perp} \to -k^{\perp}$ and $k^{\perp (m)} \to -k^{\perp (m)}$ which causes $r_{\sigma, \bar{\sigma}} \to - r_{\sigma, \bar{\sigma}}$ and  $r_{\sigma, \sigma} \to  r_{\sigma, \sigma}$. This is exactly the same result as the one we obtained by exchanging $B \to -B$. This can be understood by analyzing a rotation by angle $\pi$ around an arbitrary axis in the $x$-$y$-plane of our whole system. This rotation should leave the reflection coefficients invariant and is carried out by exchanging $k^{\perp} \to -k^{\perp}$, $k^{\perp (m)} \to -k^{\perp (m)}$ and $B \to -B$. We conclude that the parameter governing the reflection coefficients is the projection of the magnetic field onto the outward normal vector of the TI surface, cf. Fig.~\ref{fig:Figure1}.

\section{Results}
\label{sec:Results}
In Sec.~\ref{sec:Analytical results} we first find general characteristics of the Casimir force between two infinite PTI half spaces separated by vacuum with a general permittivity tensor \eqref{eq:Epsilon} and a permeability of $\mu=1$. Second in Sec.~\ref{sec:InSb Model} we calculate and analyze the Casimir force for the InSb model \eqref{eq:Epsilon InSb}.

\subsection{Analytical results in the retarded/nonretarded limits}
\label{sec:Analytical results}
In this subsection the near field (nonretarded) and the far field (retarded) limits are analyzed.
\begin{figure}
	\centering
	\includegraphics[width=1\linewidth]{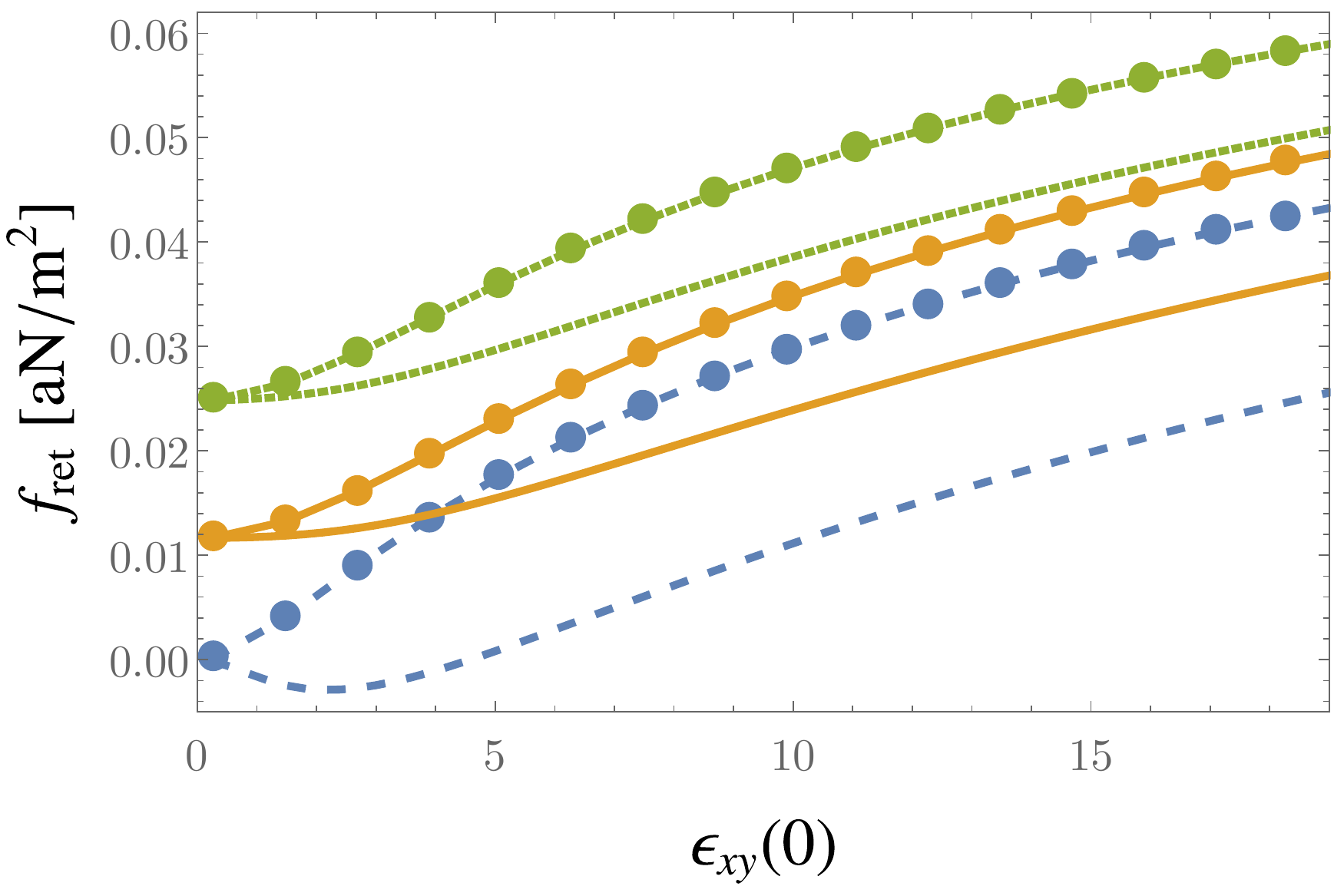}
	\caption{Numerically calculated Casimir force density for different static values of $\epsilon_{ii}$ as a function of $\epsilon_{xy}$ at a fixed gap distance of $L = 10$\,mm. The dashed lines correspond to $\epsilon_{xx}=3$ and $\epsilon_{zz} \to \infty$, the solid lines to $\epsilon_{xx}= \epsilon_{zz}= 3$ and the dotted ones to $\epsilon_{xx}=\epsilon_{zz} = 1$. Lines with (without) circles correspond to the case with $ B^+= B^-$ ($B^+ = -B^-$) in each case.}
	\label{fig:Figure2}
\end{figure} 
In the retarded limit we assume that $\omega_{\textrm{res}} L/c \gg 1$ and find that the term $\me^{-2 \xi L/c}$ restricts the frequency dependence: $0 \leq \xi \leq c/(2L) \ll \omega_{\textrm{res}}$. $\omega_{\textrm{res}}$ stands for the smallest relevant plasma or resonance frequency associated with the medium. So we can approximate $\epsilon \left( \mi \xi \right) \backsimeq \epsilon \left( 0 \right)$. We insert different static values for $\epsilon_{ij} \left( \mi \xi=0 \right)$ into Eq.~\eqref{eq:Casimir Force Final Result} and calculate the Casimir force density at a fixed distance $L = 10$ mm. We obtain for the reflection coefficients $r_{\textrm{s,p}} = r_{\textrm{p,s}} \rightarrow 0$ and $r_{\textrm{s,s}} = - r_{\textrm{p,p}} \rightarrow -1$ if $\epsilon_{xx} \rightarrow \pm \infty$ or $\epsilon_{xy} \rightarrow \pm \infty$ and for $\mi \xi \rightarrow 0$. Thus the material behaves like a perfect conductor in the retarded limit and its Casimir force is given by Eq.~\eqref{eq:Casimir Force perfect conduct}. Especially for materials with a frequency dependence of $\epsilon_{xy}$ or $\epsilon_{xx}$ similar to the Drude model we find this divergence at $\mi \xi = 0$.\\  
Fig.~\ref{fig:Figure2} shows the Casimir force for several static values of the permittivity tensor with $\epsilon_{xx}$, $\epsilon_{xy} \neq \pm \infty$ for $\mi \xi = 0$. We further distinguish between the two cases $\epsilon_{xy}^+ = \pm \epsilon_{xy}^-$ where $\epsilon_{xy}^+$, $\epsilon_{xy}^-$ are the $xy$-entry of the permittivity tensor in the left and right half space, respectively. We can see that we only find repulsive Casimir forces if $|\epsilon_{xy}| \lesssim 5$, $\epsilon_{xy}^+ = - \epsilon_{xy}^-$ and $\epsilon_{xx} = \epsilon_{zz} \cong 1 $ in the retarded limit. The reason is that  $\epsilon_{xy}$ does not only contribute to $r_{\sigma, \bar{\sigma}} $ but also to $r_{\sigma, \sigma}$. Especially for higher values of $\epsilon_{xy}$ it contributes more to $r_{\sigma, \sigma}$ than to $r_{\sigma, \bar{\sigma}}$, which results in an attractive Casimir force density. Nevertheless there is a strong dependence of $f_{\textrm{ret}}$ on the relative sign and magnitude of $\epsilon_{xy}^\pm \left( 0 \right)$. So the Casimir force of a PTI in the retarded limit $f_{\textrm{ret}}$ can be tuned by changing the external magnetic field if $\epsilon_{xx}$, $\epsilon_{xy} \neq \pm \infty$ for $\mi \xi = 0$.\\
In the nonretarded limit one cannot neglect the frequency dependence of the permittivity. Instead one can assume that $k^{\parallel} \gg \omega/c$. In this case one obtains the simplified dispersion relations $k^{\perp}_{i} \cong \mi k^{\parallel}$, $ k^{\perp (1)} \cong \mi \sqrt{\frac{\epsilon_{xx} \left( \mi \xi \right)}{\epsilon{zz} \left( \mi \xi \right)}} k^{\parallel}$ and $k^{\perp (2)} \cong \mi k^{\parallel}$. Taking only the highest order terms in $k^{\parallel}$ into account one finds $r_{\textrm{s}, \textrm{s}} \propto 1/\left( k^{\parallel} \right)^2 \cong 0 $ and $r_{\textrm{s}, \textrm{p}}, r_{\textrm{p}, \textrm{s}} \propto 1/k^{\parallel}\cong 0$. This is in accordance with the reciprocal case in which one finds $r_{\textrm{s}, \textrm{s}} \cong 0$ in the nonretarded limit as well. So the only remaining term is $r_{\textrm{p}, \textrm{p}}$ and its expression to highest order in $k^{\parallel}$ is given by 
\begin{equation}
r_{\textrm{p}, \textrm{p}} = \frac{\sqrt{\epsilon_{xx} \left( \mi \xi \right) \epsilon_{zz} \left( \mi \xi \right)} - 1}{\sqrt{\epsilon_{xx} \left( \mi \xi \right) \epsilon_{zz} \left( \mi \xi \right)} + 1}.
\end{equation}
The Casimir force in the nonretarded limit $f_{\textrm{nret}}$ reduces to a much simpler expression
\begin{equation}
\label{eq:Nonretarded}
f_{\text{nret}} = - \frac{\hbar}{8 \pi^2 L^3} \int\limits_{0}^{\infty} \dif \xi \, \textrm{Li}_3 \left[ \left( \frac{\sqrt{\epsilon_{xx} \left( \mi \xi \right) \epsilon_{zz} \left( \mi \xi \right)} - 1}{\sqrt{\epsilon_{xx} \left( \mi \xi \right) \epsilon_{zz} \left( \mi \xi \right)} +1} \right)^2 \right].
\end{equation} 
Interestingly this short distance approximation is independent of $\epsilon_{xy}$ and closely related to the case of an isotropic material where now the geometric mean $\sqrt{\epsilon_{xx} \epsilon_{zz}}$ appears in place of the isotropic permittivity. Similar results have first been found for the Casimir--Polder interaction between an atom and a uniaxial material which is isotropic on the interface plane and where the optic axis conincides with the anisotropic direction \cite{Fichet:1995, Bruch:1977, Kihara:1965}.\\
Since applying the time-reversal operator only changes the sign of $\epsilon_{xy}$, the solution in the nonretarded limit is unaffected by changing the direction of the external magnetic field $B^\pm$. This does not mean, that $f_{\textrm{nret}}$ does not change as well by tuning the external magnetic field, because $\epsilon_{ii}$ can still depend on e.g. $B^2$ as in the model of InSb. Furthermore we find that $f_{\textrm{nret}}$ is proportional to $1/L^3$ for short distances as in the reciprocal case.

\subsection{B-dependence of the Casimir force}
\label{sec:InSb Model}
After exploring the general characteristics of the Casimir force between two semi-infinite half spaces of the PTI separated by a layer of vacuum with thickness $L$, we analyze the force for the material model of InSb \eqref{eq:Epsilon InSb}. We especially want to concentrate on how the force depends on the bias magnetic fields applied to the right $\vec{B}^+$ and left $\vec{B}^-$ half spaces where $\vec{B}^\pm$ is always perpendicular to the interfaces between PTI and the vacuum. Furthermore let us note that we found in Sec.~\ref{sec:Reflection coefficients} that the sign of $r_{\sigma, \bar{\sigma}}$ depends on the projection of the applied magnetic fields onto the outward normal vector of the surfaces. So we define the projected external magnetic field of the left ($^-$) and the right ($^+$) interface as $B^\pm = \vec{B}^\pm \cdot \textbf{dA}^\pm$. Here $\textbf{dA}^\pm$ is the normal vector of the right and left half space with $\dif \vec{A}{}^+ = -\vec{e}_z$ and $\dif \vec{A}{}^- = \vec{e}_z$, cf. Fig.~\ref{fig:Figure1}. That means that with this definition the signs of $r^+_{\sigma,\bar{\sigma}}$ and $r^-_{\sigma,\bar{\sigma}}$ are the same for $B^+= B^-$ and they are opposite for $B^+ =- B^-$. So we will further differentiate between $B^+ = \pm B^-$ because we expected different results for the Casimir force in the two cases after the previous discussion.\\
The problem of realizing magnetic fields with opposite directions in the two half-spaces in an experiment has been described in Ref.~\cite{Grushin:2011_2}. One can cover the PTI with a thin ferromagnetic layer on which the influence on the Casimir force is negligible. Another way is to dope the PTI with magnetic impurities \cite{Cheng:2014}, although the material response tensor would need to be modified.
\begin{figure}
	\centering
	\includegraphics[width=1\linewidth]{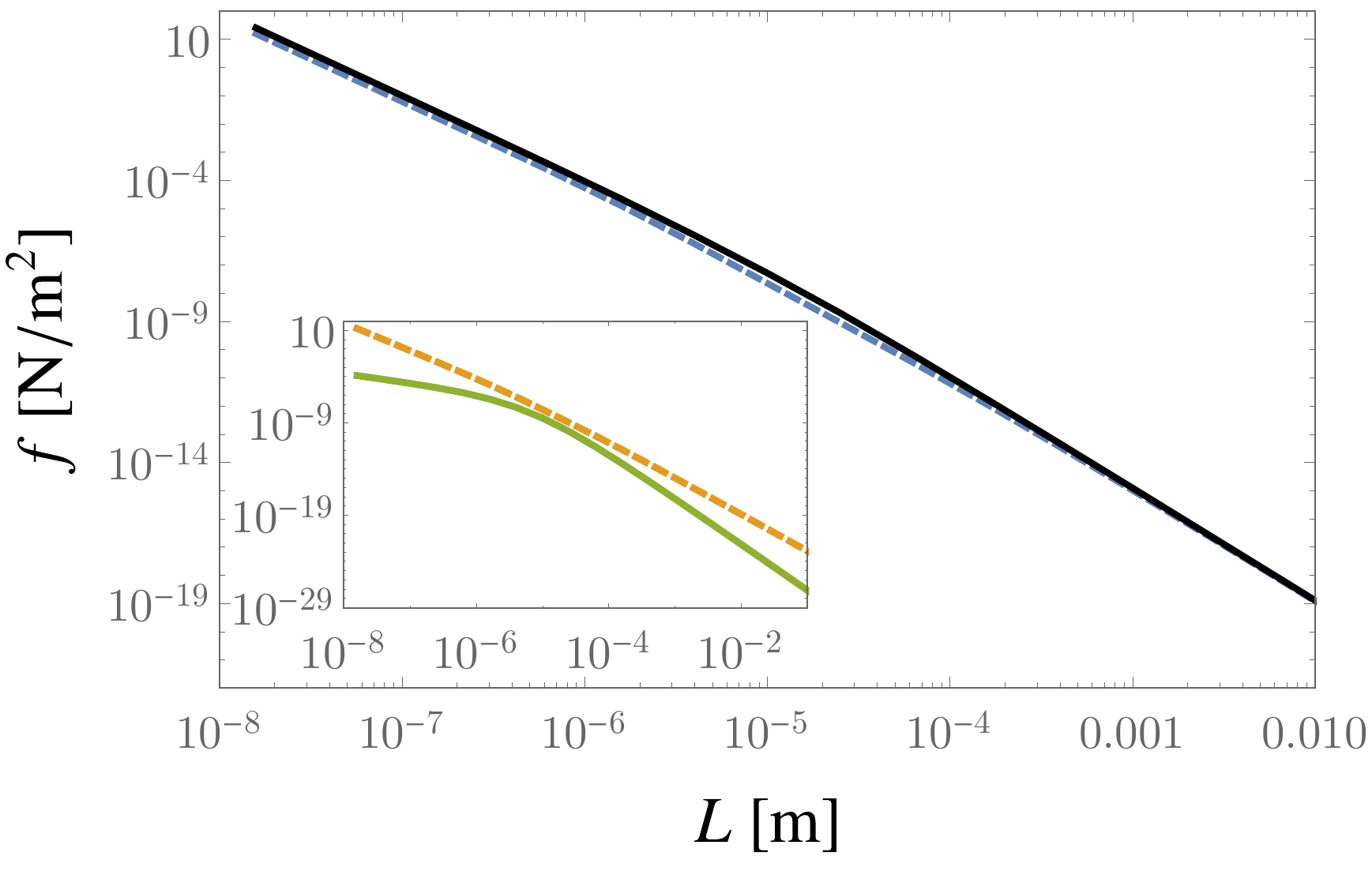}
	\caption{The Casimir force density $f$ has been calculated using Eq.~\eqref{eq:Casimir Force Final Result} for the InSb model \eqref{eq:Epsilon InSb} as a function of the gap distance $L$ in the main plot without applied magnetic field (solid line) and for the case $B^- = -B^+=20$ T (dashed line). Additionally $f$ was plotted only considering contributions from reflections with $f_\text{off}$ (solid line) or without $ f_\text{diag} $ (dashed line) change of polarization in the inset for the case $B^- = -B^+=20$ T. Since $f_\text{off}< 0$, the negative value $-f_\text{off}$ was plotted.}
	\label{fig:Figure3}
\end{figure}
Figure \ref{fig:Figure3} shows the numerical results for the Casimir force as a function of the gap distance $L$ for the two cases $B= 0$ T and $B^+ = -B^-= 20$ T. It can be seen that the influence of the bias magnetic field is small on this logarithmic scale. Nevertheless we find here that the influence vanishes for large values of $L$ because the two graphs overlap in that region but they split at intermediate distances and stay separated even in the nonretarded limit.\\
Furthermore we analyze which one of the reflection coefficients is responsible for the main contribution of the Casimir force density. The contributions of the offdiagonal (diagonal) terms $f_\textrm{off}$ $\left( f_\textrm{diag} \right)$ are depicted in Fig.~\ref{fig:Figure3}, which has been calculated by setting $r_{\textrm{s}, \textrm{s}} = r_{\textrm{p}, \textrm{p}} = 0$ $\left( r_{\textrm{s}, \textrm{p}} = r_{\textrm{p}, \textrm{s}} = 0 \right)$. Therefore only reflections where the polarization does change (does not change) are considered. As expected for the case of $B^+ = -B^-$, $f_\textrm{off}$ is purely repulsive whereas $f_\textrm{diag}$ is attractive. But since $|f_\textrm{diag}|>|f_\textrm{off}|$ the total Casimir force is attractive and mainly dominated by contributions from $r_{\textrm{s}, \textrm{s}}$, $r_{\textrm{p}, \textrm{p}}$. At intermediate distances the influence of $r_{\textrm{s}, \textrm{p}}= r_{\textrm{p}, \textrm{s}}$ is probably not negligible because the yellow and green graphs are getting close to each other at around $10^{-5}$\,m. So at this point we  expect to find differences for $B^+=\pm B^-$ because for the case $B^+= B^-$, $f_\textrm{off}$ would have the same absolute value but a different sign as for $B^+=- B^-$ as explained at the beginning of this section.\\
To analyze the $B$-dependence in greater detail, Fig.~\ref{fig:Figure4} shows the relation $f \left( B=0 \right) /f \left( B \right)$ as a function of the gap distance. As expected we find a constant value of $1$ in the retarded limit and $f \left( B=0 \right) /f \left( B \right) \neq 1$ at intermediate distances and in the nonretarded limit. Here the fraction $f \left( B=0 \right) /f \left( B \right)$ increases if $B$ increases. Interestingly there is a peak at around $10^{-5}$ m where we find a reduction of the Casimir force by a factor of up to $2.3$ for $B=20$ T. The height of that peak depends additionally on whether the two magnetic fields are pointing in the same direction or not. So at intermediate distances we can see clearly the effect of nonreciprocity. Now the question arises if that peak is only caused by $r_{\textrm{s}, \textrm{p}}$ and $r_{\textrm{p}, \textrm{s}}$.
\begin{figure}
\centering
\includegraphics[width=1\linewidth]{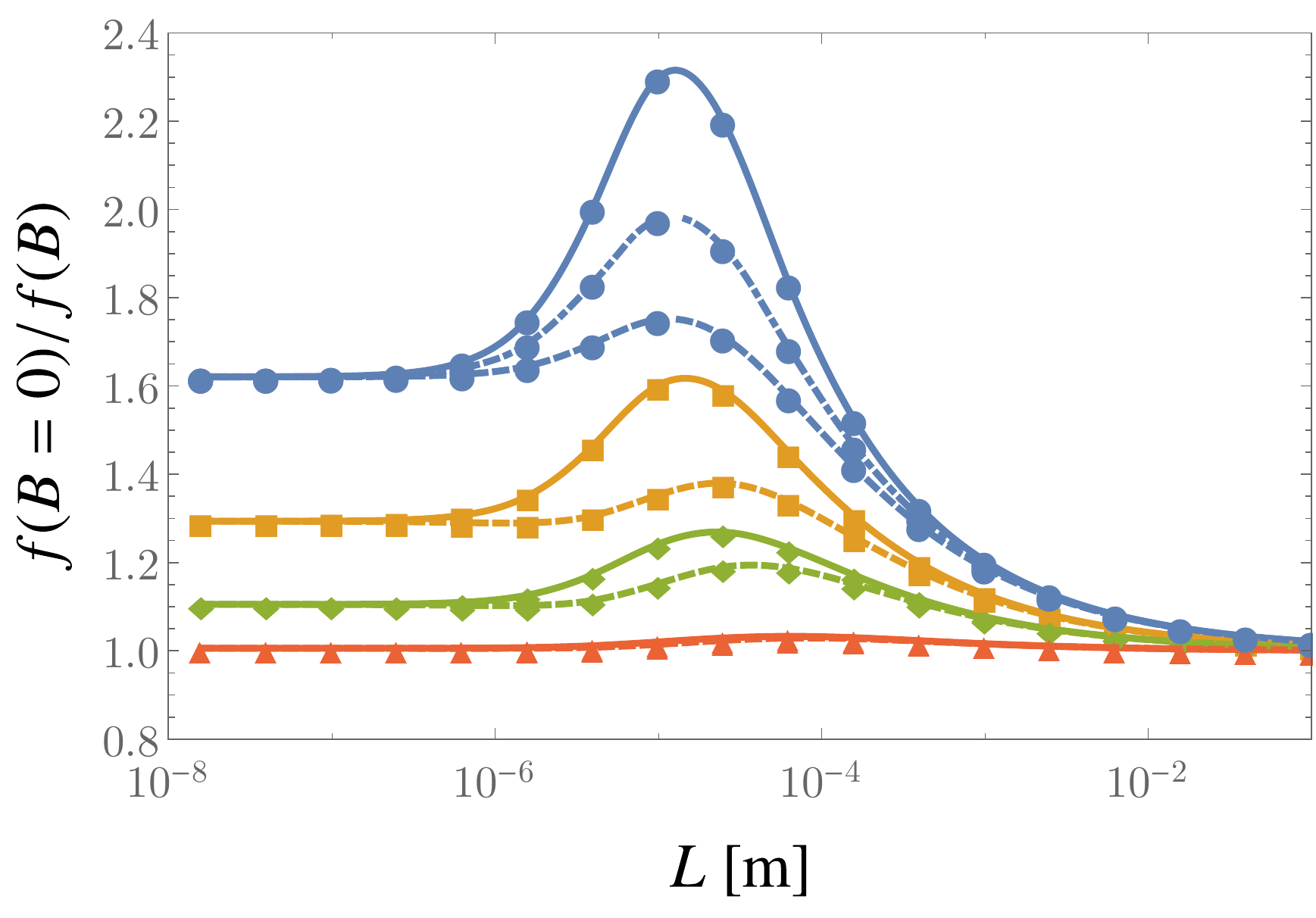}
\caption{Numerically calculated ratio between the Casimir force density with and without external magnetic field $f \left( B=0 \right) /f \left( B \right)$ as a function of the gap distance $L$ and for different magnetic fields $B= 1$ T (triangles), $5$ T (diamonds), $10$ T (squares), $20$ T (bullets). Furthermore, we distinguished between the cases with $B^+=B^-$ (dashed lines) and $B^+=-B^-$ (solid lines). For $B=20$ T we also plotted the numerical result in case of $r_{\textrm{s}, \textrm{p}}$, $r_{\textrm{p}, \textrm{s}}=0$ (dashed and dotted line) and so here the only nonvanishing contributions arise from $r_{\textrm{s}, \textrm{s}}$, $r_{\textrm{p}, \textrm{p}}$.}
\label{fig:Figure4}
\end{figure}
From the dotted and dashed lines in Fig.~\ref{fig:Figure4} one can see a sensitive region at around $L= 10^{-5}$\,m, even if $r_{\textrm{s}, \textrm{p}} = r_{\textrm{p}, \textrm{s}}=0$ and it is enhanced or reduced depending on $\textrm{sgn} \left( B^+ \right) =- \textrm{sgn} \left( B^- \right)$ (solid lines) or $\textrm{sgn} \left( B^+ \right) = \textrm{sgn} \left( B^- \right)$ (dashed lines), respectively, if $r_{\textrm{s}, \textrm{p}}$, $r_{\textrm{p}, \textrm{s}}\neq 0$ again.\\
In particular, the case $\textrm{sgn} \left( B^+ \right) = - \textrm{sgn} \left( B^- \right)$can be attributed to a repulsive Casimir force component (the force being reduced), while $\textrm{sgn} \left( B^+ \right) = \textrm{sgn} \left( B^- \right)$ corresponds to an attractive force. There is an analogy to the attractive Casimir force between two perfectly conducting or two perfectly permeable plates and the repulsive Casimir force between one perfectly conducting and one perfectly permeable plate, cf. Sec.~\ref{sec:Greens Tensor Two Surfaces}. Reference \cite{Boyer:1974} connects this behavior to the attractive Van-der-Waals force between two purely electrically polarizable particles or two purely magnetically polarizable particles and the repulsive force between one purely electrically polarizable and one purely magnetically polarizable particle. This can be extended to the attractive Casimir--Polder force of an electrically polarizable particle to a perfectly conducting wall and the repulsion of the magnetically polarizable particle from the wall \cite{Boyer:1974}. Moreover, there is a similar feature in the Casimir--Polder potential between a circularly polarized atom and an axion topological insulator, where the axion contribution stemming from a coupling between the electric and the magnetic field decreases the effect of the ordinary Casimir--Polder potential \cite{Fuchs:2017}.\\
In the nonretarded limit $f$ is reduced by a constant factor depending on the strength of the magnetic field only. That can be understood analytically by noting that $f_{\textrm{nret}}$ does not depend on $\epsilon_{xy}$ and thus is invariant under a sign change of one of the magnetic fields. Using Eq.~\eqref{eq:Nonretarded} we can plot $\tilde{f}_\textrm{nret}=f_\textrm{nret} \left( B=0 \right)/f_\textrm{nret} \left( B \right)$ as a function of the magnetic field, which is shown in Fig.~\ref{fig:Figure5} for a gap distance of $L = 10^{-9}$ m. $\tilde{f}_\text{nret}$ increases with increasing bias magnetic field until it saturates at around $40$\,T where the reduction factor has almost reached $2$. 
\begin{figure}
\centering
\includegraphics[width=1\linewidth]{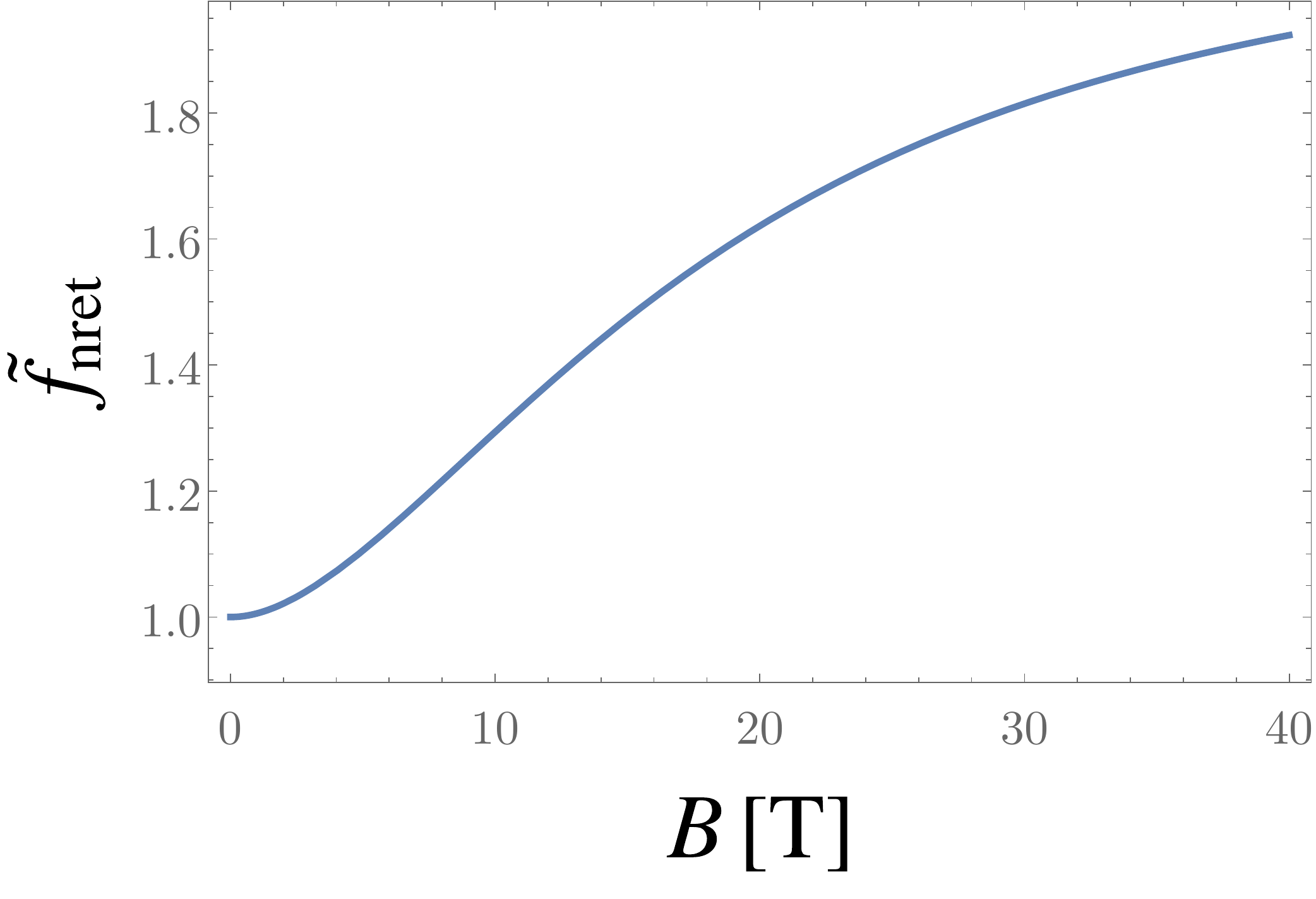}
\caption{The Casimir force in the nonretarded limit approximation $f_\textrm{nret}$ has been evaluated numerically for a fixed gap distance of $L = 1$ nm as a function of the magnetic field $B=B^\pm$. Afterwards the ratio $\tilde{f}_\textrm{nret}=f_\textrm{nret} \left( B=0 \right)/f_\textrm{nret} \left( B \right)$ has been computed and plotted. As one can see here this reduction factor $\tilde{f}_\textrm{nret}$ increases with increasing magnetic field until it saturates at around $B\cong 40$ T at $\tilde{f}_\textrm{nret} \cong 2$.}
\label{fig:Figure5}
\end{figure}
\begin{figure*}
\centering 
\subfigure{\includegraphics[width=0.2225\linewidth]{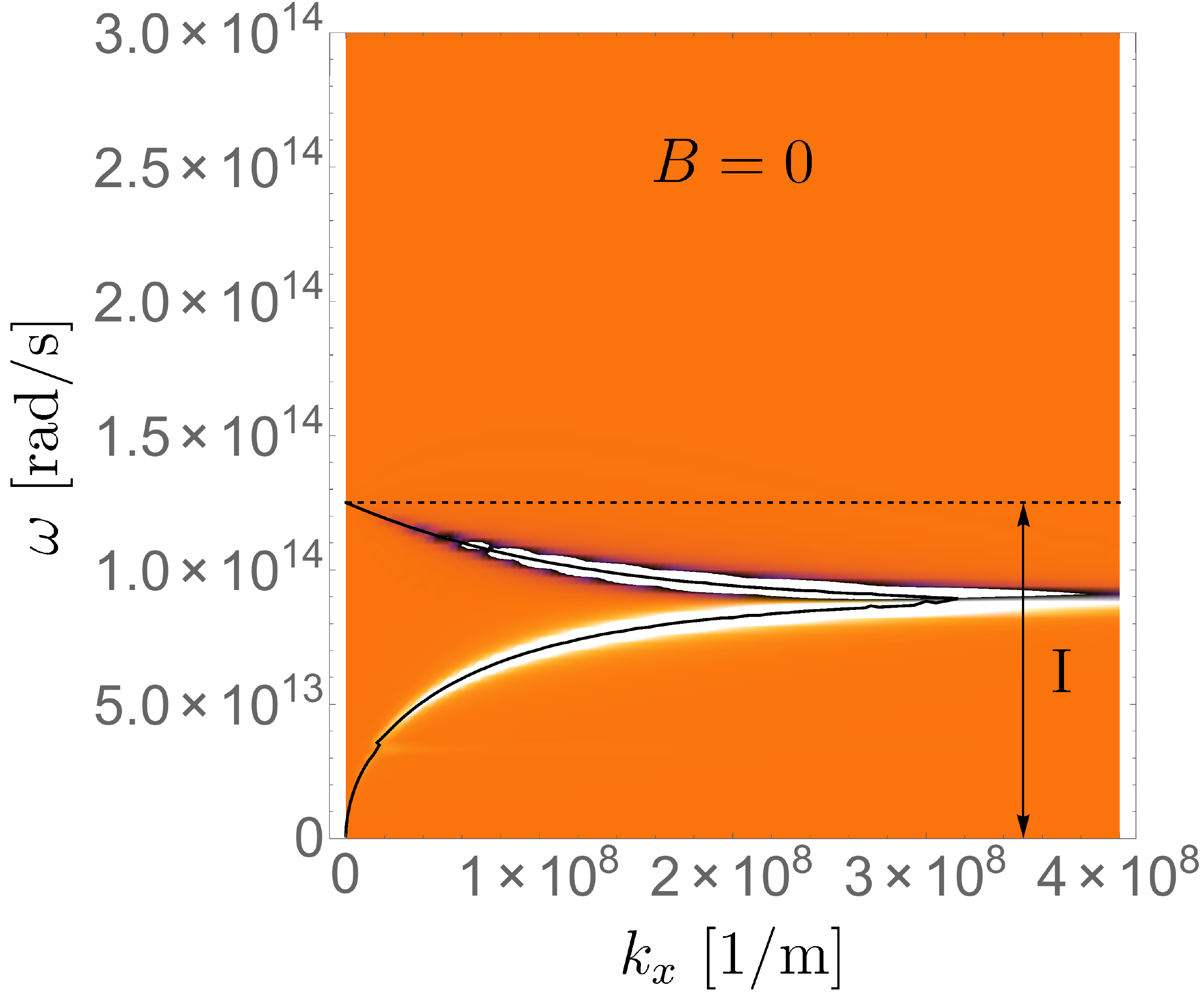}}
\subfigure{\includegraphics[width=0.163\linewidth]{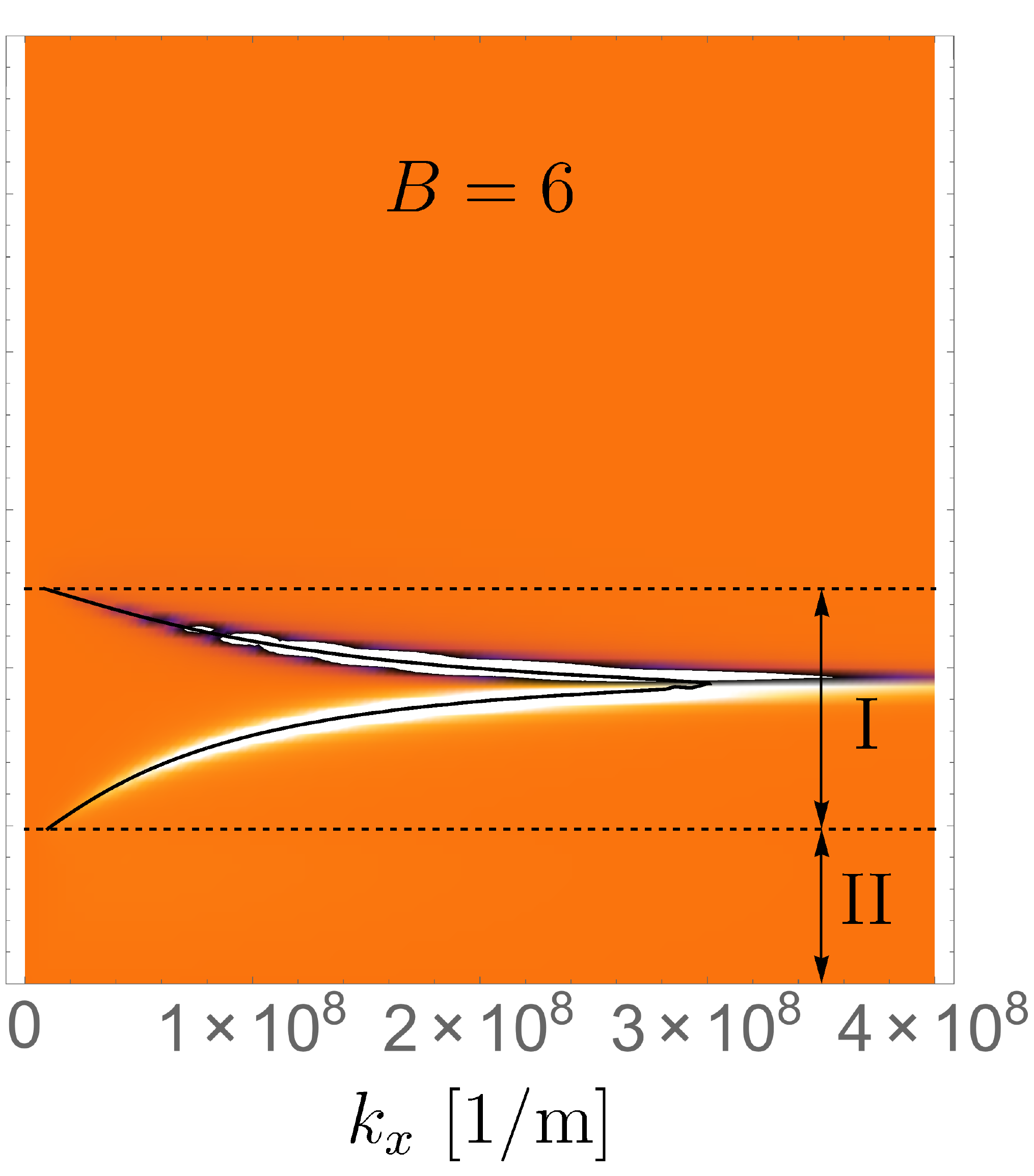}}
\subfigure{\includegraphics[width=0.163\linewidth]{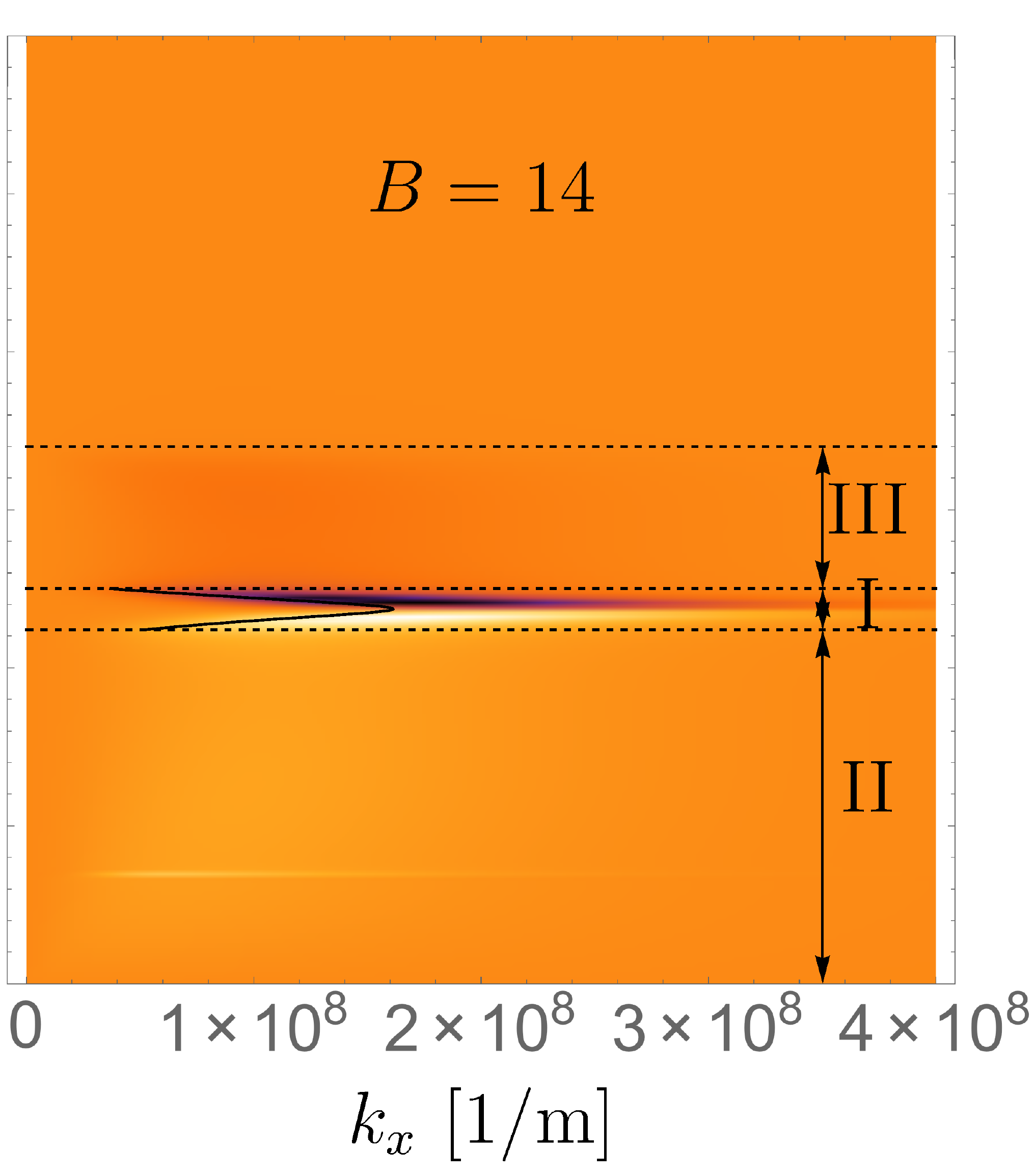}}
\subfigure{\includegraphics[width=0.163\linewidth]{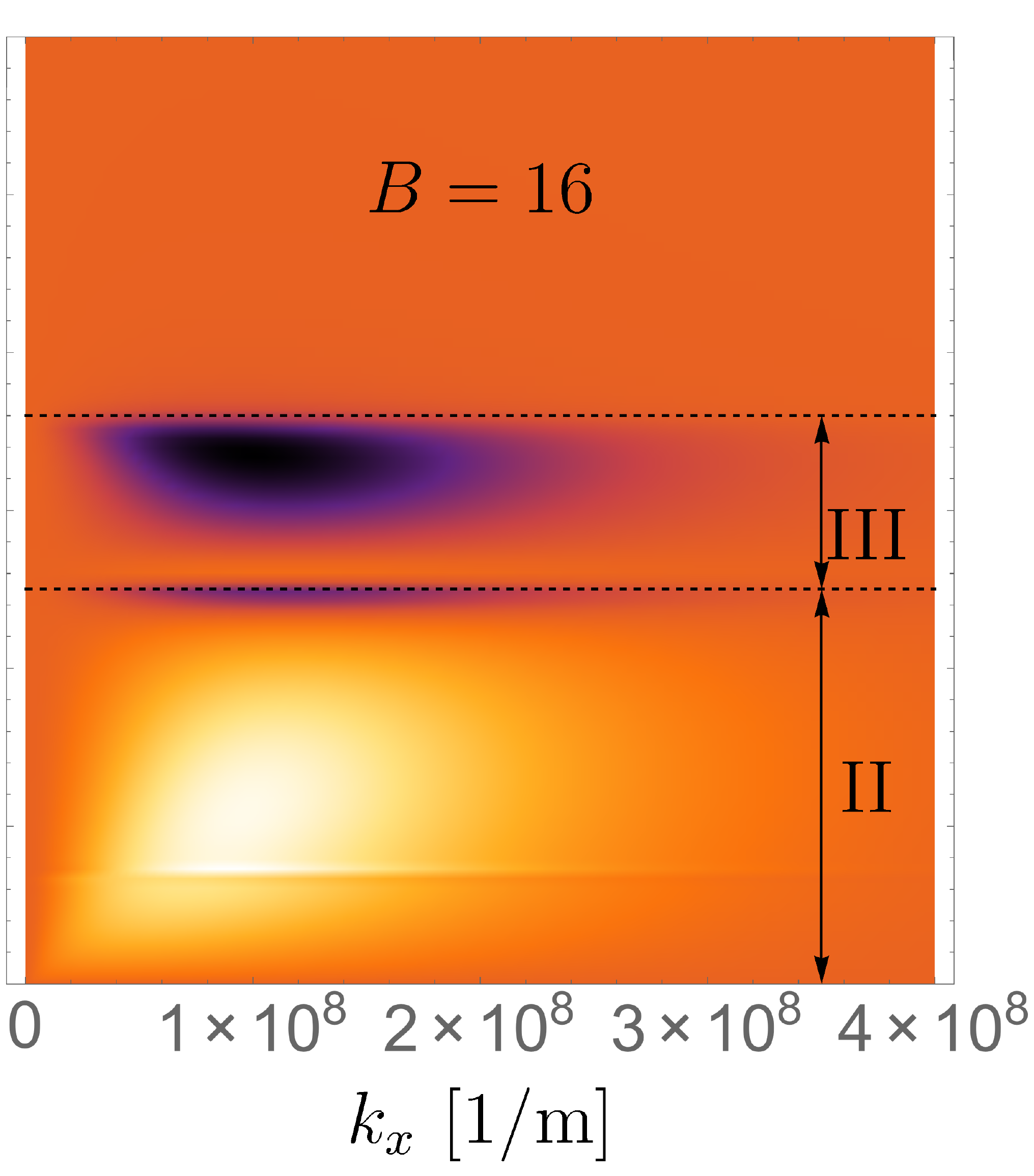}}
\subfigure{\includegraphics[width=0.163\linewidth]{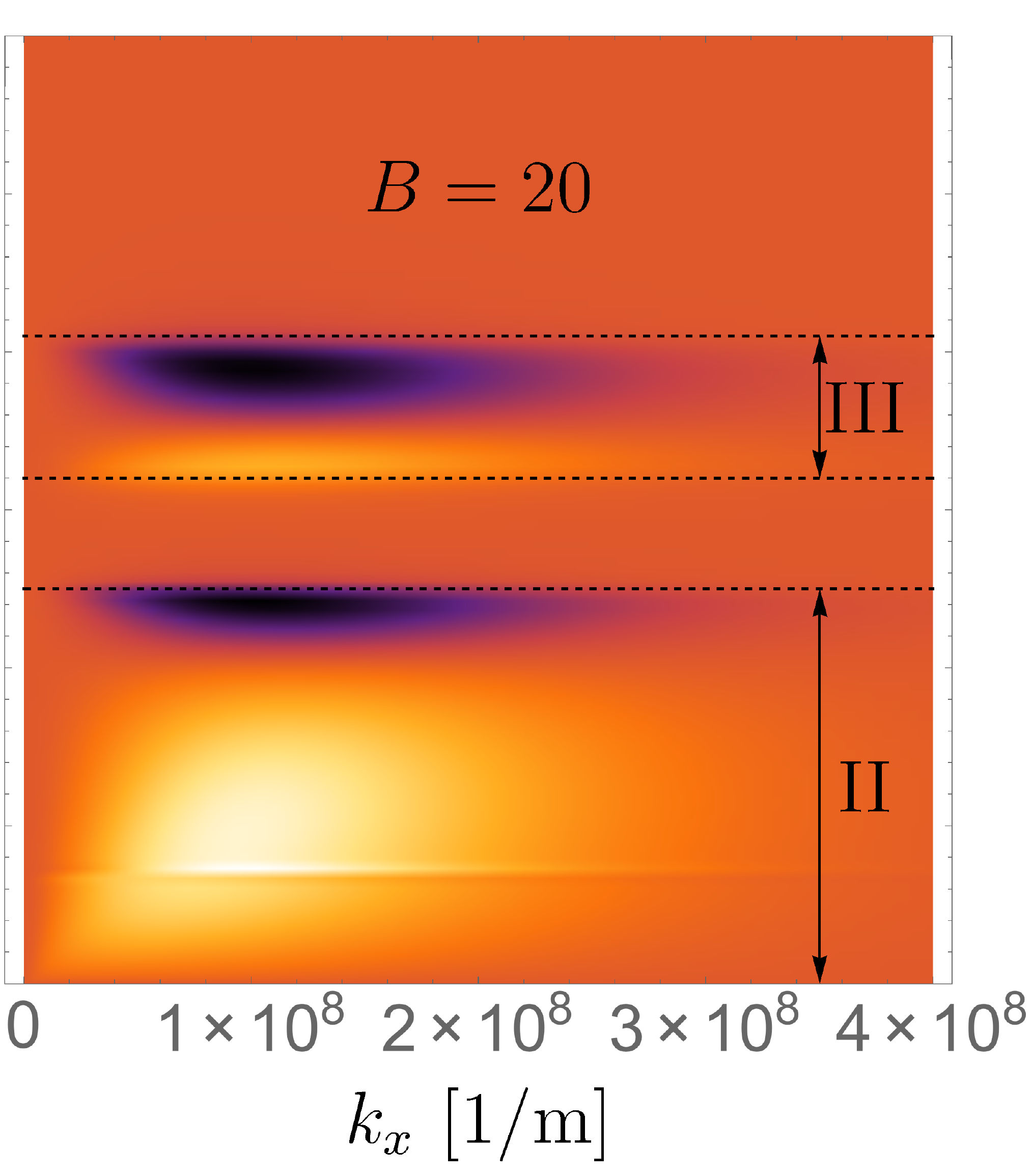}}
\subfigure{\includegraphics[width=0.08\linewidth]{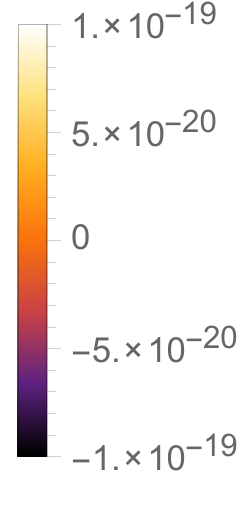}}
\caption{The Casimir force density in the nonretarded limit approximation is plotted as a function of $\omega$ and $k^x$ for different magnetic fields $B = B^\pm$ and with $L = 10$ nm, similar to Ref.~\cite{Moncada:2015}. Additionally, the dispersion relations for surface cavity modes are plotted with black solid lines in the $\omega$-$k^\parallel$-plane. Furthermore one can find different regions I, II, III, in each figure separated by black dotted lines. In I we have $\textrm{Re} \left( \epsilon_{xx} \right),\textrm{Re} \left( \epsilon_{zz} \right) < 0$ and so this is the region with surface phonon and surface plasmon polaritons. In II and III there is $\textrm{Re} \left( \epsilon_{xx} \right) \textrm{Re} \left( \epsilon_{zz} \right) < 0 $ and we can find the hyperbolic modes HMI (II) and HMII (III), respectively.}
\label{fig:Figure6}
\end{figure*}
To understand the shifting of $f$ in the nonretarded limit for an increasing magnetic field we can analyze the Casimir force as a function of $k^{\parallel}$ and $\omega$ in the nonretarded limit. The result is very similar to the one obtained for the near field heat transfer discussed in \cite{Moncada:2015}. The Casimir force density does not depend on $\epsilon_{xy}$ for $k^x \gg \omega/c$, cf. Eq.~\eqref{eq:Nonretarded}, and the only nonvanishing reflection coefficient is $r_{\textrm{p}, \textrm{p}}$. So one finds the following relation between $k^x$ and $\omega$ in the nonretarded limit for surface phonon (SPhPs) and surface plasmon polaritons (SPPs), cf. Ref.~\cite{Moncada:2015} 
\begin{equation}
k^x = \frac{1}{L} \text{ln}\left( \frac{\sqrt{\epsilon_{xx}(\omega)\epsilon_{zz}(\omega)}-1}{ \sqrt{\epsilon_{xx}(\omega)\epsilon_{zz}(\omega)} +1 } \right)
\end{equation}
if $\text{Re}(\epsilon_{xx}),\text{Re}(\epsilon_{zz})<0$. This equation can also be found by setting the denominator of the multiple reflections in the nonretarded limit $1 - r^+_{\textrm{p}, \textrm{p}} r^-_{\textrm{p}, \textrm{p}} \me^{2 \kappa^{\perp} L}$ equal to zero.\\
Furthermore the general dispersion relations simplify to
\begin{align}
\begin{array}{lll}
&k^{\perp}_1 &= \sqrt{ \epsilon_{xx} \frac{\omega^2}{c^2} - \left( k^x \right)^2 }\\
&k^{\perp}_2 &= \sqrt{ \epsilon_{xx} \frac{\omega^2}{c^2} - \frac{\epsilon_{xx}}{\epsilon_{zz}} \left( k^x \right)^2},
\end{array}
\end{align}
where the second equation can also be brought to the form
\begin{equation}
\frac{\omega^2}{c^2} = \frac{ \left( k^{\perp}_2 \right)^2}{\epsilon_{xx}} + \frac{\left( k^x \right)^2}{\epsilon_{zz}}.
\end{equation}
This describes hyperbolic modes for $\text{Re} \left( \epsilon_{xx} \right) \text{Re} \left( \epsilon_{zz} \right)< 0$, where one can further distinguish between hyperbolic modes with $\text{Re} \left( \epsilon_{xx} \right) > 0$ and $\text{Re} \left( \epsilon_{zz} \right) < 0$ (HMI) and modes with $\text{Re} \left( \epsilon_{xx} \right) < 0$ and $\text{Re} \left( \epsilon_{zz} \right) > 0$ (HMII). Interestingly these modes are therefore propagating within the material and evanescent in the vacuum and therefore frustrated internal reflections.\\      
Figure \ref{fig:Figure6} shows $f \left( \omega, k^\parallel \right)$ for different values of $B$. At $B=0$ the main contribution of the Casimir force is due to SPPs and SPhPs. By increasing the magnetic field this contribution gets less intense and is shifted to lower $k^x$-values. Furthermore the region $\textrm{Re} \left( \epsilon_{xx} \right), \textrm{Re} \left( \epsilon_{zz} \right) < 0$, where these modes are allowed, becomes smaller until it vanishes completely at around $B=16$ T.\\
On the other hand there are increasing contributions from hyperbolic modes with stronger magnetic fields, which can be found at a broad range of $\omega$- and $k^x$-values equally and which are restricted to $\textrm{Re} \left( \epsilon_{xx} \right) \text{Re} \left( \epsilon_{zz} \right) < 0$. In summary, we found that the Casimir force is dominated by SPPs and SPhPs for small magnetic fields in the nonretarded limit whereas the main contributions stem from hyperbolic modes for bigger values of $B$. So the decrease of $f$ is due to the fact that the contributions of hyperbolic modes at high fields are smaller than the ones of SPPs and SPhPs at small magnetic fields.\\
To see all the effects described in this section one needs large magnetic fields up to $20$\,T. In the following we investigate if the necessary value of the magnetic field is smaller if we used different values for the parameters (especially $n$ and $m^\star$ and so $\omega_{\textrm{p}}$ in our material model, cf. Eq.~\eqref{eq:Epsilon InSb}). Therefore we plot the Casimir force at a fixed distance $L=10$\,nm in the nonretarded limit for different magnetic fields as a function of $\omega_{\textrm{p}}$. The result is shown in Fig.~\ref{fig:Figure7}. It can be seen that there are mainly two limits for the Casimir force which are independent of $B$. One is reached at high values of $\omega_{\textrm{p}}$. In this case $\omega_{\textrm{p}} \gg \omega_{\textrm{c}}$ and there is no difference whether there is a magnetic field or not. Whereas in the other limit for small values of $\omega_{\textrm{p}}$ and so for $\omega_{\textrm{c}} \gg \omega_{\textrm{p}}$ the Casimir force has reached a minimal value, which is different from $f \left( B=0 \right)$. It depends on the value of $\omega_{\textrm{p}}$ where the transition from one limit to the other happens. The dotted vertical line shows the value of $\omega_{\textrm{p}}$ which we used in all of our calculations so far. For smaller values of $\omega_{\textrm{p}}$ a weaker magnetic field would influence the Casimir force in the same way.  
\begin{figure}
\centering
\includegraphics[width=1\linewidth]{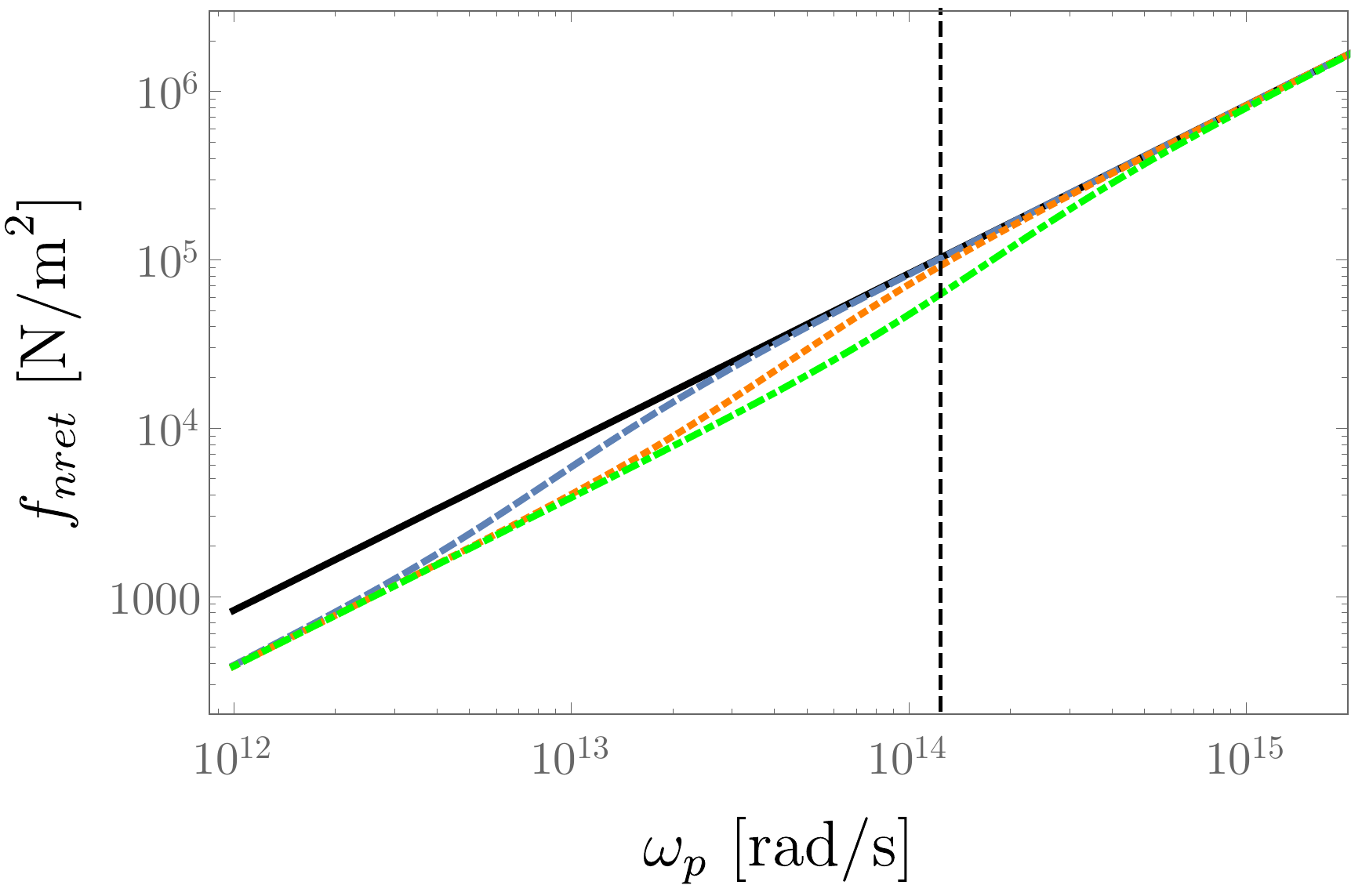}
\caption{Casimir force density in the nonretarded limit $f_\textrm{nret}$ for a fixed gap distance $L=1$ nm as a function of the plasma frequency $\omega_{\textrm{p}}$ for different fixed values of $B^\pm = 0$ T (solid line), $1 $ T (dashed line), $5$ T (dotted line) and $20$ T (dashed and dotted line).}
\label{fig:Figure7}
\end{figure}

\subsection{Casimir repulsion for a model inspired by Iron Garnet}	
The second model inspired by iron garnet, was introduced in Ref.~\cite{Silveirinha:2015} and examined in Ref.~\cite{Hanson:2016}. The respective elements of the permittivity tensor \eqref{eq:Epsilon} read
\begin{align}\label{eq:Iron Garnet Epsilon}
\begin{array}{lll}
\epsilon_{xx} = 1- \displaystyle{\frac{\omega_0 \omega_{\textrm{e}}}{\omega^2 - \omega^2_0}}\\
\epsilon_{zz} = 1 \\
\epsilon_{xy} = -\epsilon_{yx} =  \displaystyle{\frac{\omega \omega_{\textrm{e}}}{ \omega^2 - \omega^2_0}}.
\end{array}
\end{align}
In this model $|\omega_0|$ is the resonance frequency and $\omega_{\textrm{e}}$ is the resonance strength. As shown in Ref.~\cite{Silveirinha:2015}, it is sufficient to claim $\omega_{\textrm{e}} \omega_0 > 0$. One finds from Eq.~\eqref{eq:Iron Garnet Epsilon} that by changing the sign of $\omega_{\textrm{e}}$ and $\omega_0$ one can achieve $\epsilon_{xy}  \to -\epsilon_{xy}$, whereas the diagonal elements $\epsilon_{ii}$ do not change their signs.\\
So far there is no experimental evidence for a real material with such a permittivity. The exact $B$-field dependence of $\Greektens{$\epsilon$}$ is unknown. Nevertheless we consider it as an alternative hypothetical model which can be compared to the one of InSb and which shows a different aspect concerning the Casimir force. 
\begin{figure}
\centering
\includegraphics[width=1\linewidth]{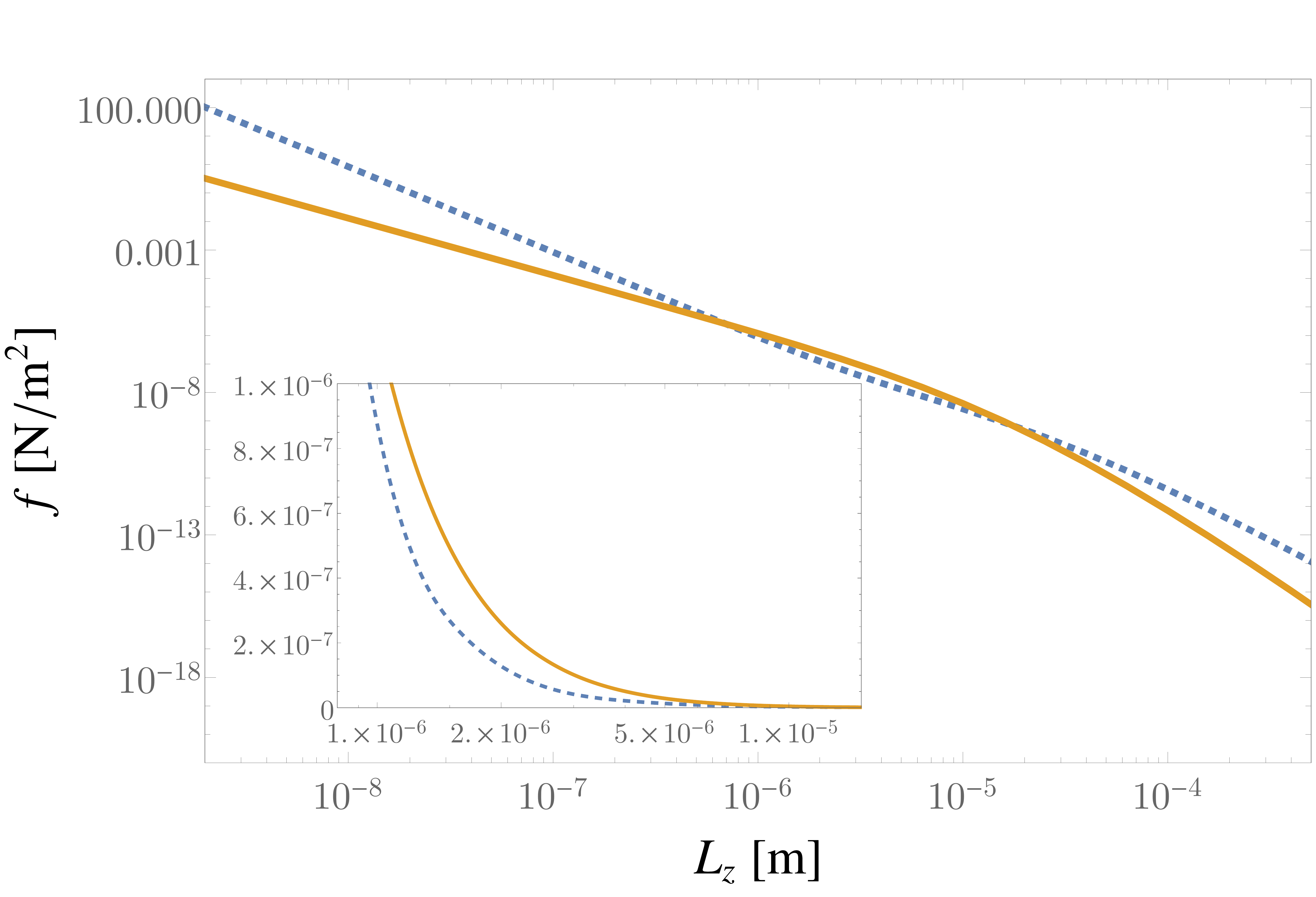}
\caption{The different contributions from reflection with $\left(- f_\textrm{off} \right)$ (solid line) or without $\left( f_\textrm{diag} \right)$ (dashed line) a change of polarization at the interface are plotted separately for the parameters $\omega_{\textrm{e}} = \pm 7.7 \cdot 10^{12}$\,$2 \pi/$s and $\omega_0 =\pm 1.3 \cdot 10^{10}$\,$2 \pi/$s and additionally $\omega_0 \omega_{\textrm{e}} < 0$ in both half spaces. Since $f_\textrm{off}<0$ we plotted $-f_\textrm{off}$. There is a region where $|f_\textrm{off}|>|f_\textrm{diag}|$ (see inset) and so that is where we expect to find a total repulsive Casimir force, cf. Fig.~\ref{fig:Figure9}.}
\label{fig:Figure8}
\end{figure}
\begin{figure}
\centering	 	
\includegraphics[width=1\linewidth]{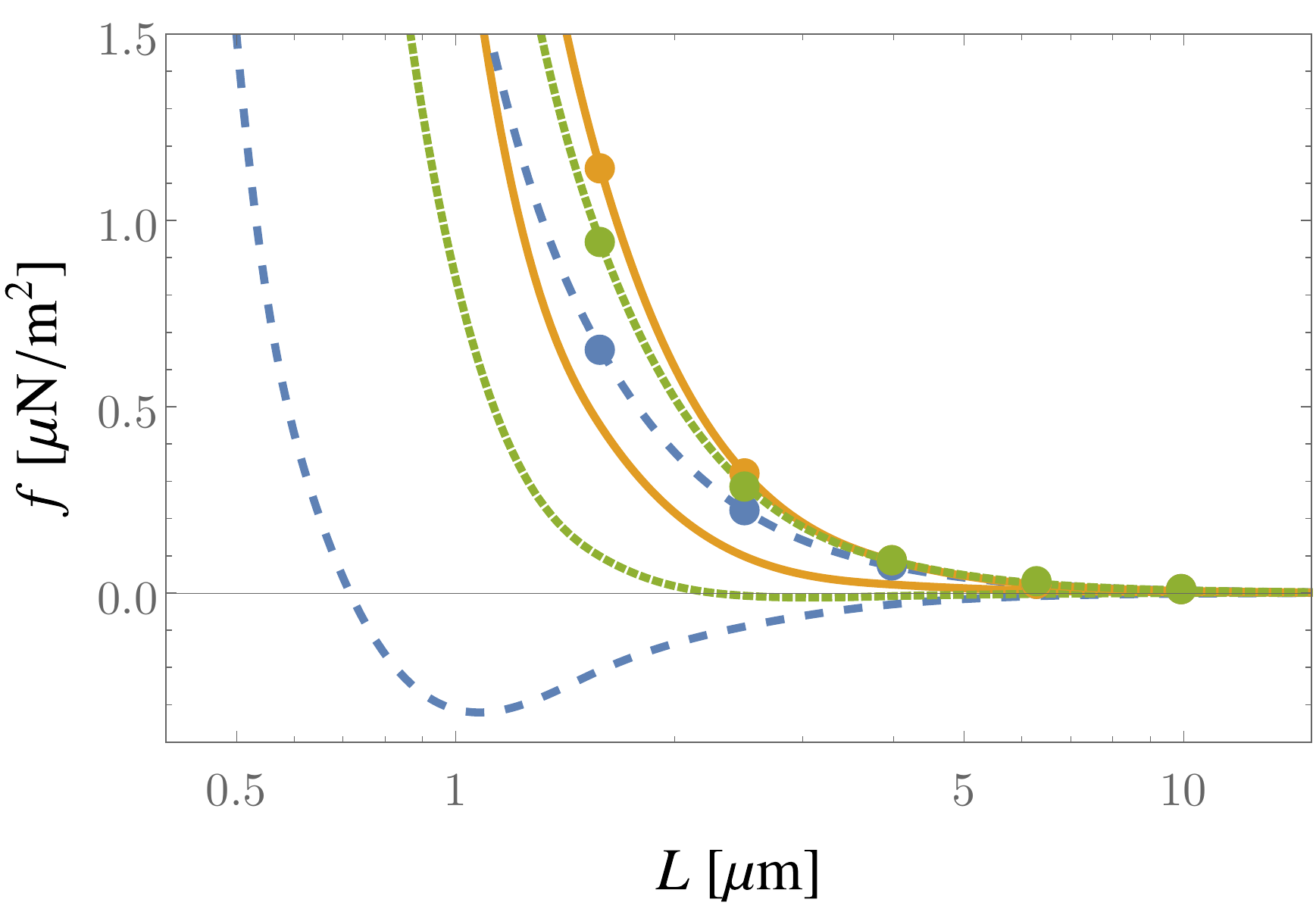}
\caption{The Casimir force density of a PTI for the model similar to iron garnet as a function of the gap distance. We used in both half spaces $\omega_{\textrm{e}} = \pm 7.7 \cdot 10^{12}$\,$2 \pi/$s and $\omega_0 =\pm 1.3\cdot10^{m_2}$\,$2 \pi/$s and different values for $m_2= 10$ (dashed), $11$ (dotted) and $12$ (solid). Additionally, we distinguished between the two cases where first $\omega_0,$ $\omega_{\textrm{e}} > 0 $ in both half spaces (without circles) and second $\omega_0$, $\omega_{\textrm{e}} < 0 $ in only one of the half spaces and $\omega_0,$ $\omega_e > 0 $ in the other one (with circles).}
\label{fig:Figure9}
\end{figure}
Similar to Sec.~\ref{sec:InSb Model} we calculate the Casimir force numerically by using the permittivity values similar to iron garnet \eqref{eq:Iron Garnet Epsilon} and insert them into Eq.~\eqref{eq:Casimir Force Final Result}. First we have a look at the different contributions only from reflections with ($f_\textrm{off}$) or without ($f_\textrm{diag}$) change of polarizations, for the case where $\omega_0$, $\omega_{\textrm{e}} < 0$ in both of the half spaces. The result is shown in Fig.~\ref{fig:Figure8}. We find that in the retarded and nonretarded limit $|f_\textrm{off}|< f_\textrm{diag}$ for InSb whereas the reflections with a change of polarization dominate at intermediate distances. So we expect the force to be repulsive in that region. This assumption is confirmed in Fig.~\ref{fig:Figure9}. So there is a region at around $2\cdot 10^{-7} \, \text{m}<L<10^{-4}$ m where we can switch between a repulsive and an attractive force simply by changing the sign of $\omega_0$ and $\omega_{\textrm{e}}$ in one of the half spaces.

\section{Conclusion}
We have derived a general expression for the Casimir force density between two nonreciprocal semi-infinite half spaces. This derivation is based on an extension of the theory of macroscopic quantum electrodynamics for nonreciprocal material.\\
This general expression is applied to a photonic topological insulator with a permittivity tensor with offdiagonal elements. First we have derived the reflection coefficients and investigated the Casimir force based on the general expression of the permittivity tensor. The reflection coefficients with a polarization flip at the interface change signs by switching from a positive to a negative magnetic field in only one of the half spaces. Whereas the material behaves like a perfect conductor in the retarded limit if $\epsilon_{xx} \rightarrow \pm \infty$ or $\epsilon_{xy} \rightarrow \pm \infty$ for $\mi \xi \rightarrow 0$, the reflection coefficient for parallel polarization dominates in the nonretarded limit and we give an analytical result for the Casimir force.\\
We then applied the Casimir force formalism to the material model of InSb. We found a dependence of the Casimir force on the magnitude of the magnetic field at small distances. The polarization changing reflection coefficients were shown to be responsible for a repulsive force only if the signs of the magnetic fields in the two half spaces differ, whereas the other components cause an attractive force regardless of the sign. Nevertheless, this model never allows for a repulsive net force. Only a reduction of the magnitude of the Casimir force can be observed if an external magnetic field is applied. This can be explained by studying the force in the nonretarded limit. In this case the Casimir force is dominated by surface phonon and surface plasmon polaritons at small magnetic fields, whereas they are outperformed by hyperbolic modes at larger magnetic fields.
The ratio between the Casimir force without applied field and the force with magnetic field shows a maximum at a critical distance of $L=10^{-5}$ m where the Casimir force for the InSb model is reduced by a factor of about 2 at $10$\,T. Additionally in this regime of intermediate distances the magnitude of the Casimir force strongly depends on the relative sign of the magnetic fields in the two half spaces.\\
Finally, the Casimir force was studied for a model inspired by iron garnet. In this case the impact of the polarization changing reflection coefficients is stronger than the terms which do not change the polarization at intermediate distances. This makes it possible to achieve repulsive Casimir forces.

\section{acknowledgments}
We would like to thank Diego Dalvit and Robert Bennett for discussions. This work was supported by the German Research Foundation (DFG, Grants BU 1803/3-1 and GRK 2079/1). S.F. and F.L. are grateful for the hospitality at the Chemistry Department of the University of British Columbia, where this work was mostly done. S.Y.B is grateful for support by the Freiburg Institute of Advanced Studies.


%

\end{document}